\documentclass[a4paper,11pt]{article}
\pdfoutput=1
\usepackage{jcappub} 
\usepackage[T1]{fontenc} 

\usepackage{graphicx}
\usepackage{multirow}
\usepackage{subfigure}
\usepackage{booktabs}
\newcommand{\be}{\begin{equation}}
\newcommand{\ee}{\end{equation}}
\newcommand{\bea}{\begin{eqnarray}}
\newcommand{\eea}{\end{eqnarray}}

\title{\boldmath Prospects of Future CMB Anisotropy Probes for Primordial Black Holes}

\author[a,b]{Junsong Cang,}
\author[a]{Yu Gao,}
\author[c,d]{Yin-Zhe Ma}

\affiliation[a]{Key Laboratory of Particle Astrophysics, Institute of High Energy Physics, Chinese Academy of Sciences, Beijing, 100049, China}
\affiliation[b]{School of Physical Sciences, University of Chinese Academy of Sciences, Beijing, 100049, China}
\affiliation[c]{School of Chemistry and Physics, University of KwaZulu-Natal, Westville Campus, Private Bag X54001, Durban, 4000, South Africa}
\affiliation[d]{NAOC–UKZN Computational Astrophysics Centre (NUCAC), University of KwaZulu-Natal, Durban, 4000, South Africa}

\emailAdd{cangjs@ihep.ac.cn}
\emailAdd{gaoyu@ihep.ac.cn}
\emailAdd{ma@ukzn.ac.za}

\abstract{Cascade of particles injected as Hawking Radiation from Primordial Black Holes (PBH) can potentially change the cosmic recombination history by ionizing and heating the intergalactic medium, which results in altering the anisotropy spectra of the Cosmic Microwave Background (CMB). In this paper, we study the expected sensitivity of several future CMB experiments in constraining the abundance of PBHs distributed in $10^{15}\sim10^{17}$ g mass window according to four mass functions: the monochromatic, log-normal, power-law and critical collapse models. Our result shows that future experiments, such as CMB-S4 and PICO, can improve current {\it{Planck}} bounds by about two orders of magnitude. Most regions in PBH parameter space that are allowed by current CMB data, including monochromatically distributed PBHs with mass heavier than $3.8 \times 10^{16}$ grams, can be tested by upcoming missions with high significance.
}
\begin{document}
\maketitle
\flushbottom

\section{Introduction}

Conditions in the early universe may have led to the formation of Primordial Black Holes ~\cite{Novikov1967The,10.1093/mnras/152.1.75,1975ApJ...201....1C,ca03510k,1979A&A....80..104N}. 
There are many mechanisms through which PBH can be created ~\cite{Novikov1967The,10.1093/mnras/152.1.75,1979A&A....80..104N,HAWKING198235,1975ApJ...201....1C,ca03510k,HAWKING199036,PhysRevD.78.063515,Liu:2019lul,Gao:2018pvq,Khlopov:2008qy,Belotsky:2018wph}, including the primordial inhomogeneities ~\cite{10.1093/mnras/152.1.75} and collapse of domain wall bubbles \cite{Liu:2019lul}, etc. 
Such black holes serve as an essential probe for the physics of the early universe and could manifest itself as cold dark matter (CDM) ~\cite{article,Carr:2016drx,Ali-Haimoud:2019khd}.
The allowed masses for PBH can span over a wide range from $10^{-38}{\rm{M}}_{\odot}$ to $10^5{\rm{M}}_{\odot}$ or higher \cite{Carr:2009jm}, depending on the time of PBH formation after the Big Bang. 
PBHs in various mass ranges~\cite{Carr:2009jm,Poulin:2016anj,Ali-Haimoud:2016mbv,Ali-Haimoud:2019khd,Montero-Camacho:2019jte,Arbey:2019vqx,Ukwatta:2015tza} can be probed by 
gravitational wave~\cite{Chen:2019xse}, 
lensing effect~\cite{PhysRevLett.111.181302}, emission from accretion of matter~\cite{Ricotti_2008,Tashiro:2012qe,Ali-Haimoud:2016mbv}, 
CMB spectral distortions due to dissipation of small scale scalar perturbations ~\cite{Kohri:2014lza,Nakama:2017xvq,2012ApJ...758...76C},
etc, see Ref.\cite{Carr:2020xqk} for a recent review.

Hawking radiation is suitable to study the PBHs in the mass range of $[10^{15},10^{17}]$ g~\cite{Mack:2008nv,PhysRevD.78.023004,Clark:2016nst,Poulter:2019ooo}. 
In this mass window the emissions are dominantly $e^{\pm},\ \gamma,\ \nu$. $e^{\pm}$ and $\gamma$ are electromagnetically interactive and deposit part of their kinetic energies into ionization of the intergalactic medium (IGM) after recombination, which increases the number density of free electrons and their scattering rates with CMB photons, leaving distinctive signatures on CMB anisotropy spectrum~\cite{Padmanabhan:2005es}.
Fraction of the energy radiated from PBH can also be deposited to heat up IGM temperature or cause deviations in CMB energy spectra~\cite{Slatyer:2015kla}.

PBHs in this mass window can be further constrained by extragalactic photon background (EGB) ~\cite{Carr:2009jm,Carr:2016drx}, diffuse supernova neutrino background ~\cite{Dasgupta:2019cae,Wang:2020uvi},
galactic gamma-ray measurement~\cite{Laha:2019ssq,Dasgupta:2019cae,Laha:2020ivk,DeRocco:2019fjq,Prantzos:2010wi,Kierans:2019pkh}, etc.
Currently the dominant constraint comes from the 21cm measurement by EDGES ~\cite{Clark:2018ghm,Bowman:2018yin}.
However these constraints can suffer from huge uncertainties on background astrophysical models, most noticeably the galactic dark matter density profile and cosmic-ray propagation in our galaxy~\cite{Laha:2019ssq}. 
In comparison, bounds set by CMB anisotropy are more robust as most of the calculations involved are linear and based on well understood physics~\cite{Padmanabhan:2005es}.

In this work we focus on the prospective PBH constraints from future CMB experiments.
Over the past few years several experiments has been proposed to measure CMB anisotropy with higher precision than {\it{Planck}},
including satellite missions such as COrE ~\cite{DiValentino:2016foa,Delabrouille:2017rct}, LiteBIRD ~\cite{LiteBIRD} and PICO ~\cite{Hanany:2019wrm,Hanany:2019lle,Young:2018aby}, 
as well as ground-based experiments such as 
AdvACTPol \cite{Calabrese:2014gwa}, AliCPT ~\cite{Li:2017lat,Li:2018rwc}, Simons Array \cite{Arnold:2014qym,Creminelli:2015oda}, SPT-3G \cite{Anderson:2018mry} and CMB-S4 ~\cite{Abazajian:2019eic,Abazajian:2016yjj}.
These experiments are expected to further strengthen current CMB bound on PBH set by {\it{Planck}} data, 
which has already been studied in a number of references ~\cite{Poulin:2016anj,Poulter:2019ooo,Clark:2016nst,Stocker:2018avm,Acharya:2020jbv}.


The structure of this paper is as follows: 
In Section \ref{Evaporation_Physics} we review particle and energy emissions of Hawking radiation. 
Section \ref{IGM_interaction} discusses the energy injection and deposition process for PBHs with monochromatic mass distribution, 
we then generalize our treatment to include PBHs with extended mass distributions in Section \ref{Mass_Functions}.
Section \ref{HyRec_and_CAMB} discusses the recombination history in the presence of PBH injection and the relevant impact on the CMB anisotropy spectrum. 
We outline our forecasting procedure in Section \ref{Forecast_Setup} and present our results in Section \ref{Results}. 
Section \ref{Discussion} concludes this paper. 

\section{Hawking Radiation} \label{Evaporation_Physics}
The Hawking evaporation from a Schwarzschild black hole with mass $M$ is described by a thermal emission spectrum at a temperature~\cite{1975CMaPh..43..199H}
\begin{equation} \label{T}
T_{\rm{PBH}}=
\frac{1}
{8\pi G M}
=
1.06\ {\rm{TeV}} \times
\left(
\frac{10^{10}\ {\rm{g}}}
{M}
\right),
\end{equation}
where $G$ denotes the gravitational constant. 
The number of particles emitted per energy and time interval is given by~\cite{1975CMaPh..43..199H,MacGibbon:1990zk},
\begin{equation} \label{dN/dEdt}
\frac{{\rm{d}}N^{\alpha}}
{{\rm{d}} \varepsilon {\rm{d}}t}
=
\frac{g^{\alpha}}
{2 \pi}
\frac{\Gamma_{\rm{s}}}
{{\rm{e}}^{ \varepsilon /T_{\rm{PBH}}}-(-1)^{2s}},
\end{equation}
Here the superscript ${\alpha}$ labels the particle species, $g^{\alpha}$ is the degree of freedom for the particle,
$s$ and $\varepsilon$ denote the spin and total energy of the emitted particle respectively, $\Gamma_{\rm{s}}$ is the absorption probability~\cite{Carr:2009jm},
\be
\label{Gamma_Relativistic}
\Gamma_{s}(M,\varepsilon)
=
\frac{\sigma_{s}(\varepsilon,M)}
{\pi}
\varepsilon^2 ,
\ee
where $\sigma_{\rm{s}}$ is the absorption cross section.
At high energies where $\varepsilon \gg T_{\rm{PBH}}$ or equivalently $\varepsilon/T_{\rm{PBH}} \to \infty$, $\sigma_{\rm{s}}$ for all particle species approaches the geometric optics limit~\cite{MacGibbon:1990zk,Carr:2009jm},
\begin{equation} \label{Sigma_g}
\sigma_{\rm{g}}
=27 \pi G^2 M^2
.
\end{equation}

At lower energies $\sigma_{\rm{s}}$ is a function of $\varepsilon$, $M$ and $s$ and is solved numerically~\cite{MacGibbon:1990zk}. 
Here we adopt the $\sigma_{\rm{s}}$ given by Ref.~\cite{Ukwatta:2015iba} for $0 \le \varepsilon/T_{\rm{PBH}} \le 21$, and assume $\sigma_{\rm{s}}=\sigma_{\rm{g}}$ when $\varepsilon/T_{\rm{PBH}} > 21$, 
which gives a reasonable approximation for the full $\sigma_{\rm{s}}$ function.

The mass loss rate caused by Hawking radiation of species ${\alpha}$ can be computed by,
\begin{equation} \label{M_dot}
\dot{M^{\alpha}}
=
- \int {\rm{d}} \varepsilon \cdot
\varepsilon
\frac{{\rm{d}} N^{\alpha}}
{{\rm{d}} \varepsilon {\rm{d}}t}
,
\end{equation}
which yields~\cite{1991PhRvD..44..376M},
\begin{equation}
\dot{M^{\alpha}}
=
-5.34 \times 10^{25} g^{\alpha} \omega^{{\alpha}}(M) 
\left(\frac{M}{\rm{g}}\right)^{-2}
\ 
{\rm g}/{\rm s}
,
\end{equation}
In the relativistic limit~\cite{1991PhRvD..44..376M}, which corresponds to very light PBHs ($M \ll 10^{11} {\rm{g}}$),
\begin{equation}
\renewcommand{\arraystretch}{1.3}
\begin{tabular}{ccc}
$\omega_{s=0} = 0.267,$ & $\omega_{s=1} = 0.060,$ & $\omega_{s=3/2} = 0.022,$\\
$\omega_{s=2} = 0.007,$ & $\omega^\nu_{s=1/2} = 0.147,$ & $\omega^{e\pm}_{s=1/2} = 0.142.$
\end{tabular}
\label{equ:bh_radiation_frac}
\end{equation}

While thermal distribution has a high-energy tail, emissions of particles heavier than the BH temperature is exponentially suppressed.
As a result, $\omega^{{\alpha}}$ for massive particle species decreases as black hole mass increases (lower $T_{\rm{PBH}}$).
For electron and positron emissions, one can approximate $\omega^{\rm{e}^{\pm}}$ by \cite{Stocker:2018avm,Poulter:2019ooo},
\begin{equation}
\omega^{\rm{e}^{\pm}}=0.142\ {\rm{exp}} \left( -\frac{M}{9.4 \times 10^{16} {\rm{g}}} \right)
.
\end{equation}

In the PBH mass range we consider, the only particles that are emitted at appreciable amount are $\gamma,\ \nu,\ {\rm{e}}^{\pm}$. 
As neutrinos are not electromagnetically interactive, hereafter we will restrict our discussions to the emission of $\gamma$ and ${\rm{e}}^{\pm}$. 
It should be noted that although $\sum_{\alpha} \dot{M^{\alpha}}$ is nonzero, the resulting mass loss $\Delta M$ is negligible compared to $M$ in our mass range. 
Even for the most radiant PBH with $M=10^{15} {\rm{g}}$, the fractional mass loss throughout the age of the universe is only at about $-\Delta M /M \sim 2\%$. We will thus assume that PBH mass remains constant across the history of the universe.

\section{Energy injection and deposition} \label{IGM_interaction}
In this section we will focus on PBHs with monochromatic ($\delta$) mass distribution, for which the energy injection rate per unit volume by emission of particle of species ${\alpha}$ is given by,
\begin{equation}
\left(
\frac{{\rm{d}}E}
{{{\rm{d}}V}{\rm{d}}t}
\right)_{\rm{INJ}}^{\delta,\alpha}
=
-
\dot{M} ^{\alpha}
n_{\rm{PBH}}.
\end{equation}
Here $\alpha=[e^{\pm},\gamma]$,
$n_{\rm{PBH}}$ is the number density of PBH.
\begin{equation} \label{n_PBH}
\begin{aligned}
n_{\rm{PBH}} &= 
f_{\rm{PBH}}
\frac{\Omega_{\rm{DM}} \rho_{\rm{cr}} (1+z)^3}
{M}
\end{aligned}
\end{equation}
where $f_{\rm{PBH}} \equiv {\Omega_{\rm{PBH}}}/ {\Omega_{\rm{DM}}}$ is the fraction of DM that consist of PBHs, $\rho_{\rm cr}$ is the critical density of the universe today.

Summing over contributions from all particle species gives the overall injection rate,
\begin{equation} \label{Injection_Rate}
\begin{aligned}
\left(
\frac{{\rm{d}}E}
{{{\rm{d}}V}{\rm{d}}t}
\right)^{\delta}_{\rm{INJ}}
 =
\sum_{\alpha}
\left(
\frac{{\rm{d}}E}
{{{\rm{d}}V}{\rm{d}}t}
\right)_{\rm{INJ}}^{\delta,\alpha} 
 =
3.67 \times 10^{25} F_{\rm{PBH}} \Omega_{\rm{DM}} \rho_{\rm{cr}} (1+z)^3
\ 
{\rm{s}}^{-1}
,
\end{aligned}
\end{equation}
where,
\be
F_{\rm{PBH}}
\equiv
f_{\rm{PBH}}
\frac{
\sum_{\alpha} \left[ {g^{\alpha}} \omega^{\alpha} \right]}
{0.688}
\left(
\frac{M}
{{\rm{g}}}
\right)^{-3},
\label{Def_F}
\ee
in which $0.688$ is the relativistic value of $\sum_{\alpha} [{g^{\alpha}} \omega^{\alpha}]$. 

After injection from PBH, $e^{\pm}$ and $\gamma$ deposit their energies through a series of interactions with IGM and CMB photons. For distortions in CMB anisotropy, the most important deposition channels are: ionization (ion.) and excitation (exc.) of hydrogen and the heating of IGM (heat.). 
The rate of energy deposition into each channel ($\rm{c}$) relates to the injection rate through a deposition efficiency $f^{\alpha}_{\rm{c}}$,
\begin{equation}
\left(
\frac{{\rm{d}}E}
{{{\rm{d}}V}{\rm{d}}t}
\right)_{\rm{DEP,c}}^{\delta,\alpha}
 =
f^{\alpha}_c
\times
\left(
\frac{{\rm{d}}E}
{{{\rm{d}}V}{\rm{d}}t}
\right)_{\rm{INJ}}^{\delta,\alpha}
.
\end{equation}

From the first principle, $f^{\alpha}_c$ can be constructed from the differential deposition efficiency given in Ref.~\cite{Slatyer:2015kla}: $T^{\alpha}_{{\rm{c}}, ijk} \equiv T^{\alpha}_c(z_{i},\varepsilon_{j},z_{k}) {\rm{d}}\log(1+z)$. 
For a primary particle of species ${\alpha}$ injected at $z_{k}$ with energy $\varepsilon_{j}$, this quantity gives the fraction of $\varepsilon_{j}$ deposited into channel ${\rm{c}}$ at redshift $z_{i}$, during the time step corresponding to ${\rm{d}}\log (1+z)$. 
For a general energy injection spectrum, the relevant deposition efficiency is given by~\cite{Slatyer:2015kla,Cang:2020exa},
\be \label{Dep_Eff_1}
\begin{aligned}
&f^{\alpha}_c (z_i)
\approx
\frac{H(z_i)(1+z_i)^3}
{\sum_j \varepsilon_j I^{\alpha} (z_i,\varepsilon_j) {\rm{d}} \varepsilon_j}
\sum_k
\frac{
1
}
{
(1+z_k)^3 H(z_k)
}
\sum_j \varepsilon_j I^{\alpha}(z_k,\varepsilon_j) T^{\alpha}_{\textrm{c}, ijk} {\rm{d}} \varepsilon_j
,
\end{aligned}
\ee
where 
$I^{\alpha}(z,\varepsilon)$ describes the particle injection rate per unit $\varepsilon$ and unit volume,
\begin{equation} \label{I_Fun_Def}
I^{\alpha}(z,\varepsilon)
\equiv
\frac{{\rm{d}}N^{\alpha}}
{{\rm{d}} \varepsilon {\rm{d}} t {\rm{d}} V} (z,\varepsilon)
=
\frac{{\rm{d}}N^{\alpha}}
{{\rm{d}} \varepsilon {\rm{d}}t}
\times
n_{\rm{PBH}}.
\end{equation}

The overall PBH deposition rate is given by summing over all particle species,
\begin{equation}
\begin{aligned}
\left(
\frac{{\rm{d}}E}
{{{\rm{d}}V}{\rm{d}}t}
\right)^{\delta}_{\rm{DEP,c}} 
& =
\sum_{\alpha}
\left(
\frac{{\rm{d}}E}
{{{\rm{d}}V}{\rm{d}}t}
\right)_{\rm{DEP,c}}^{\delta,\alpha} 
 =
f_{\rm{c}} \times
\left(
\frac{{\rm{d}}E}
{{{\rm{d}}V}{\rm{d}}t}
\right)^{\delta}_{\rm{INJ}} 
\end{aligned}
\end{equation}
where $f_{\rm{c}}$ is an overall deposition efficiency,
\begin{equation}
\begin{aligned}
f_{\rm{c}}
& \equiv
\frac{
\sum_{\alpha}
g^{\alpha}
\omega^{\alpha}
f^{\alpha}_{\rm{c}}
}
{
\sum_{\alpha}
g^{\alpha}
\omega^{\alpha}
}
 =
\frac{4 \omega^{{\rm{e}}^{\pm}} f^{{\rm{e}}^{\pm}}_{\rm{c}}+2 \omega^{\gamma} f^{\gamma}_{\rm{c}}}
{4 \omega^{{\rm{e}}^{\pm}}+2 \omega^{\gamma}}.
\end{aligned}
\end{equation}

\section{PBH Mass Functions} \label{Mass_Functions}
Many of the existing PBH constraints in the literatures~\cite{Clark:2016nst,Acharya:2020jbv,Laha:2019ssq,DeRocco:2019fjq} normally include monochromatic mass distribution. 
Although such distribution model is theoretically viable~\cite{Pi:2017gih}, many other PBH formation scenarios tend to predict extended mass distributions~\cite{PhysRevD.47.4244,Carr:2016hva,Clesse:2015wea,Kannike:2017bxn,Yokoyama:1998xd,Niemeyer:1999ak,Carr:2017jsz,Green:2016xgy,Bellomo:2017zsr}, which can generally be described by a PBH mass function,
\be
\begin{aligned}
\Psi(M) 
&\equiv
\frac{1}
{f_{\rm{PBH}}\rho_{\rm{DM}}}
\frac{
{\rm{d}}{\rho_{\rm{PBH}}(M)}
}
{{\rm{d}}M}.
\end{aligned}
\ee

A monochromatic mass spectrum is simply,
\bea
\Psi(M')
=
\delta_{\rm D}(M-M').
\eea

We will also consider three types of extended PBH mass functions:
\begin{itemize} 
\item Log-normal model ~\cite{PhysRevD.47.4244,Carr:2017jsz,Green:2016xgy}
\end{itemize}
\bea
\Psi(M)
=
\frac{1}
{\sqrt{2 \pi} \sigma M}
{\rm{exp}}
\left(
-
\frac{({\rm{log}}[M/M_c])^2}
{2 \sigma^2}
\right).
\eea

\begin{itemize}
\item Critical Collapse model~\cite{Carr:2017jsz,Yokoyama:1998xd,Niemeyer:1999ak}
\end{itemize}
\bea
\Psi(M)
\propto
M^{2.85} {\rm{exp}} (-(M/M_c)^{2.85}).
\eea

\begin{itemize}
\item Power-law model~\cite{Carr:2017jsz} 
\end{itemize}
\bea
\Psi(M)
\propto
M^{\gamma-1}
\ 
(M_{\rm{min}} < M < M_{\rm{max}}).
\label{PWL_MF}
\eea

In general, the log-normal model can provide a good approximation for many other distributions and is theoretically motivated in multiple PBH formation scenarios~\cite{Carr:2017jsz,PhysRevD.47.4244,Kannike:2017bxn,Green:2016xgy,Bellomo:2017zsr}.
The Critical Collapse model is possible if the density fluctuation leading to PBHs formation has a $\delta$ function power spectrum, whereas PBHs produced from scale-invariant power spectrum can have a power-law distribution~\cite{Carr:2017jsz}.
Note that the power law index $\gamma$ in Eq(\ref{PWL_MF}) is given by $\gamma=-2 \omega /(1+\omega)$,
where $\omega$ is the equation of state parameter at the epoch during which PBHs formed.
Here we only consider post-inflation PBH formation in which $\gamma$ ranges in $[-1,1]$.
In this context ($\gamma < 1$), the power-law $\Psi(M)$ peaks at $M_{\rm{min}}$ and becomes more concentrated as $\gamma$ decreases, so $\gamma$ serves as a good indicator of the distribution width of $\Psi(M)$.
Special treatment is required for $\gamma=0$, which corresponds to PBHs formation during matter-dominated phase of the universe. 
We will ignore this scenario and refer interested readers to Ref.\cite{Carr:2017edp} for detailed analysis.

\begin{figure}[t] 
\centering
\includegraphics[width=12cm]{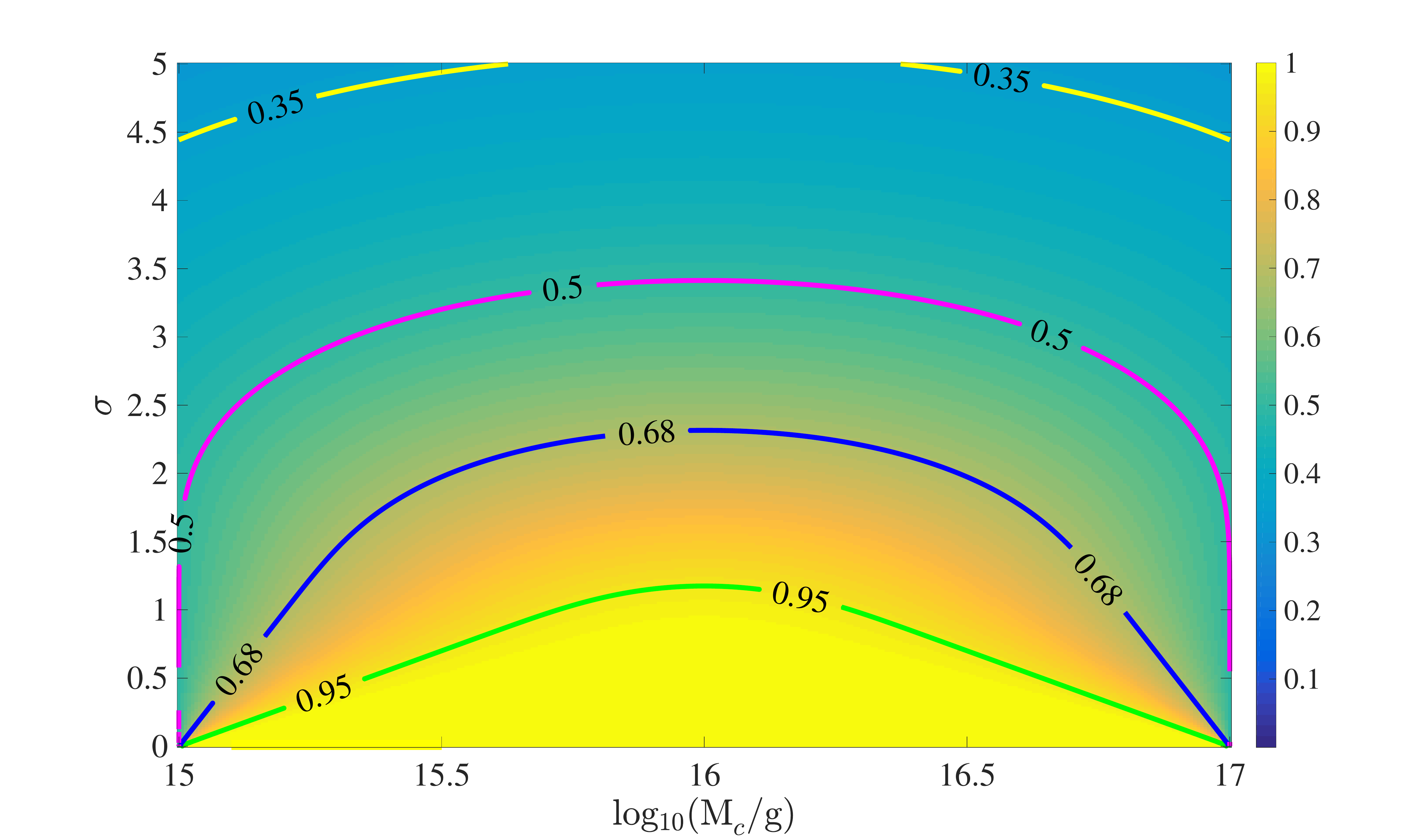}
\caption{Fraction of PBHs density contained in $10^{15} {\rm{g}} \sim 10^{17} {\rm{g}}$ mass range for log-normal model, given by $\int^{10^{17}{\rm{g}}}_{10^{15}{\rm{g}}} {\rm{d}}M \Psi $.
}
\label{Log_Normal_68_Region}
\end{figure}

With these generalizations, the total deposition rate now takes the form,
\begin{equation} \label{Extended_Treatment}
\left(
\frac{{\rm{d}}E}
{{{\rm{d}}V}{\rm{d}}t}
\right)_{\rm{DEP,c}} 
=
\int^{10^{17}{\rm{g}}}_{10^{15}{\rm{g}}}
{\rm{d}} M \cdot \Psi(M)
\left(
\frac{{\rm{d}}E}
{{{\rm{d}}V}{\rm{d}}t}
(M)
\right)^{\delta}_{\rm{DEP,c}}.
\end{equation}
Note that our mass functions are normalized for $M \in [0,\infty]$,
whereas Eq(\ref{Extended_Treatment}) only account for holes with $M \in [10^{15} {\rm{g}},10^{17} {\rm{g}}]$,
this could overestimate PBH abundance if a significant amount of PBH density were distributed outside this mass window.
The critical collapse mass function is very sharp and can be well approximated by a semi-monochromatic log-normal distribution with $\sigma = 0.26$ \cite{Carr:2017jsz}, so it is safe from this problem as long as $M_c$ is not very close to our cut-off masses.
This issue can also be avoided in power-law model by enforcing $M_{\rm{min}} \ge 10^{15} {\rm{g}}$ and $M_{\rm{max}} \le 10^{17} {\rm{g}}$.
However as shown in Fig.\ref{Log_Normal_68_Region}, the log-normal mass function is susceptible to this problem, especially for very large $\sigma$ values. 
Therefore for log-normal distributions with $\int^{10^{17}{\rm{g}}}_{10^{15}{\rm{g}}} {\rm{d}}M \Psi \le 68\%$, our constraints should be considered as conservative.

\section{Impact on Recombination} \label{HyRec_and_CAMB}
In presence of PBH injection, the recombination equations for ionization fraction $x_{\rm{e}}$ and matter temperature $T_{\rm{IGM}}$ becomes~\cite{Chen:2003gz,Liu:2016cnk}

\begin{equation}
\frac{{\rm{d}}x_{\rm{e}}}
{{\rm{d}} t}
 =
\left(
\frac{{\rm{d}}x_{\rm{e}}}
{{\rm{d}} t}
\right)_0 
+
I_{\rm{ion}} + I_{\rm{exc}},
\end{equation}

\begin{equation}
\frac{{\rm{d}}T_{\rm{IGM}}}
{{\rm{d}} t}
=
\left(
\frac{{\rm{d}}T_{\rm{IGM}}}
{{\rm{d}} t}
\right)_0 
+ 
\frac{2 Q_{\rm{heat}}}{3(1+f_{\rm{He}}+x_{\rm{e})}}.
\end{equation}
Here the standard evolution equations without additional energy injection are labeled with the subscript 0~\cite{Liu:2016cnk,AliHaimoud:2010dx}, $f_{\rm{He}}$ is the helium fraction by number of nuclei. $I_{\rm{ion}}$ and $I_{\rm{exc}}$ describe the hydrogen ionization from ground state and excited $n=2$ state respectively,
\bea
I_{\rm{ion}}(z)&=&\frac{1}{n_{\rm{H}}(z)E_i}\left(\frac{{\rm{d}}E}{{\rm{d}}V{\rm{d}}t}\right)_{{\rm{DEP,ion}}}, \\
I_{\rm{exc}}(z)&=&\frac{1-C}{n_{\rm{H}}(z)E_\alpha}\left(\frac{{\rm{d}}E}{{\rm{d}}V{\rm{d}}t}\right)_{{\rm{DEP,exc}}},
\eea
where $E_i=13.6{\rm{eV}}$, $E_{\alpha}=10.2{\rm{eV}}$. $n_{\rm{H}}$ is the number density of hydrogen nuclei, $C$ is the probability for a $n=2$ hydrogen atom to transit back to $n=1$ ground state before getting ionized~\cite{Liu:2016cnk}.
$Q_{\rm{heat}}$ is the IGM heating rate induced by PBH,
\begin{equation}
Q_{\rm{heat}}(z)=
\frac{1}{n_{\rm{H}}(z)}
\left(
\frac{{\rm{d}}E}
{{{\rm{d}}V}{\rm{d}}t}
\right)_{\rm{DEP,heat}} .
\end{equation}
It should be noted that, after the heating by PBHs' radiation, the raised IGM temperature means that collisional ionization might become more manifest at certain redshifts~\cite{Chluba:2006bc,Chluba:2010fy,Chluba:2015lpa}. This issue is beyond the scope of this paper and is not included in our numerical process.

CMB can provide a powerful probe for the ionizing effect of PBH, which increases the ionization fraction and thereby enhancing scattering rate between CMB photons and free electrons.
Such effect can leave fingerprints on CMB anisotropy by damping small scale temperature and polarization correlations while shifting peak locations for polarization anisotropy spectra \cite{Adams:1998nr,Padmanabhan:2005es}. The PBH heating term can increase gas temperature and be constrained through experiments that measure the 21cm emission line from neutral hydrogen~\cite{Mack:2008nv,Clark:2018ghm,Lopez-Honorez:2016sur,Mellema:2012ht,Trott:2019lap}.

\section{Forecast Likelihood} \label{Forecast_Setup}
At this forecast we use temperature and E-mode polarization correlations. 
Foreground contamination is assumed to have been fully removed and we only use frequency channels between 89 and 160 GHz, which roughly corresponds to the frequency range where CMB signal is expected to have the highest intensity compared to foreground components such as synchrotron and dust~\cite{Ade:2018sbj}. 
Our mock data $\hat C_{\ell}$ can be expressed as a combination of fiducial CMB signal $\bar C_{\ell}$ and experimental noise $N_{\ell}$,
\bea
\hat C_{\ell} = \bar C_{\ell} + N_{\ell},
\eea
\begin{equation}
\label{Fiducial_CL} 
\bar C_{\ell} \equiv 
\begin{bmatrix}
\bar C_{\ell}^{TT} & \bar C_{\ell}^{TE} \\
\bar C_{\ell}^{TE} & \bar C_{\ell}^{EE}\\
\end{bmatrix}
, \ 
N_{\ell} \equiv 
\begin{bmatrix}
N_{\ell}^{TT} & 0 \\
0 & N_{\ell}^{EE}\\
\end{bmatrix},
\end{equation}
here $\bar C_{\ell}s$ are generated by CAMB~\cite{Lewis:1999bs} codes for a flat $\rm{\Lambda CDM}$ cosmology set by $\it{Planck}$ 2018 data~\cite{Aghanim:2018eyx}: $\Omega_{\rm{b}}h^2=0.02242$, $\Omega_{\rm{c}}h^2=0.11933$, $100\theta_{{\rm{MC}}}=1.04101$, $\tau=0.0561$, ${\rm{log}}(10^{10}A_{\rm{s}})=3.047$, $n_{\rm{s}}=0.9665$.
$N_{\ell}$ is given by~\cite{Errard:2015cxa},
\bea \label{eq:Nlv}
N^{{\rm{EE}}}_{\ell}&=&\left[\sum_{\nu} \frac{1}{N^{\rm{EE}}_{\ell,\nu}}\right]^{-1},\ N^{{\rm{TT}}}_{\ell} = N^{{\rm{EE}}}_{\ell}/2,
\eea
\bea
N^{{\rm{EE}}}_{\ell,\nu}=\delta P^2_{\nu}\ {\exp}\left[\ell (\ell +1)\frac{\theta^2_{\rm{FWHM},\nu}}{8\ {\ln}2}\right],
\label{eq:noise_par}
\eea
where $\delta P_{\nu}$ denotes the instrumental white noise level for frequency $\nu$, measured in $\mu {\rm{k}}\cdot  {\rm{rad}}$. $\theta_{\rm{FWHM,\nu}}$ is the Full-Width at Half-Minimum beam size in radians. 
For an experiment with a sky coverage of $f_{\rm{sky}}$, 
the effective $\chi^2$ for the likelihood used in our analysis can be written as~\cite{Hamimeche:2008ai},
\bea
\label{Likelihood}
\begin{aligned}
& \chi^2_{\rm{eff}}
\equiv
-2{\ln}\mathcal{L} 
 = f_{\rm sky} \sum_{\ell}(2\ell +1)\left[{\rm{Tr}}({\hat{C}_{\ell}}C_{\ell}^{-1}) -{\rm{log}}|{\hat{C}_{\ell}}C_{\ell}^{-1}|-2\right],
\end{aligned}
\eea
where $C_{\ell}$, given by
\begin{equation}
\label{Signal_CL} 
C_{\ell} \equiv 
\begin{bmatrix}
C_{\ell}^{TT} & C_{\ell}^{TE} \\
C_{\ell}^{TE} & C_{\ell}^{EE}\\
\end{bmatrix},
\end{equation}
is calculated by CAMB for each sampled point in the parameter space for a ``${\rm{\Lambda CDM}}+{\rm{PBH}}$" cosmology.
Specifications for experiments considered in our forecast can be found in Table \ref{Tab_ExpSpecs}.

\section{Results} \label{Results}
\begin{figure}[t] 
\centering
\includegraphics[width=12cm]{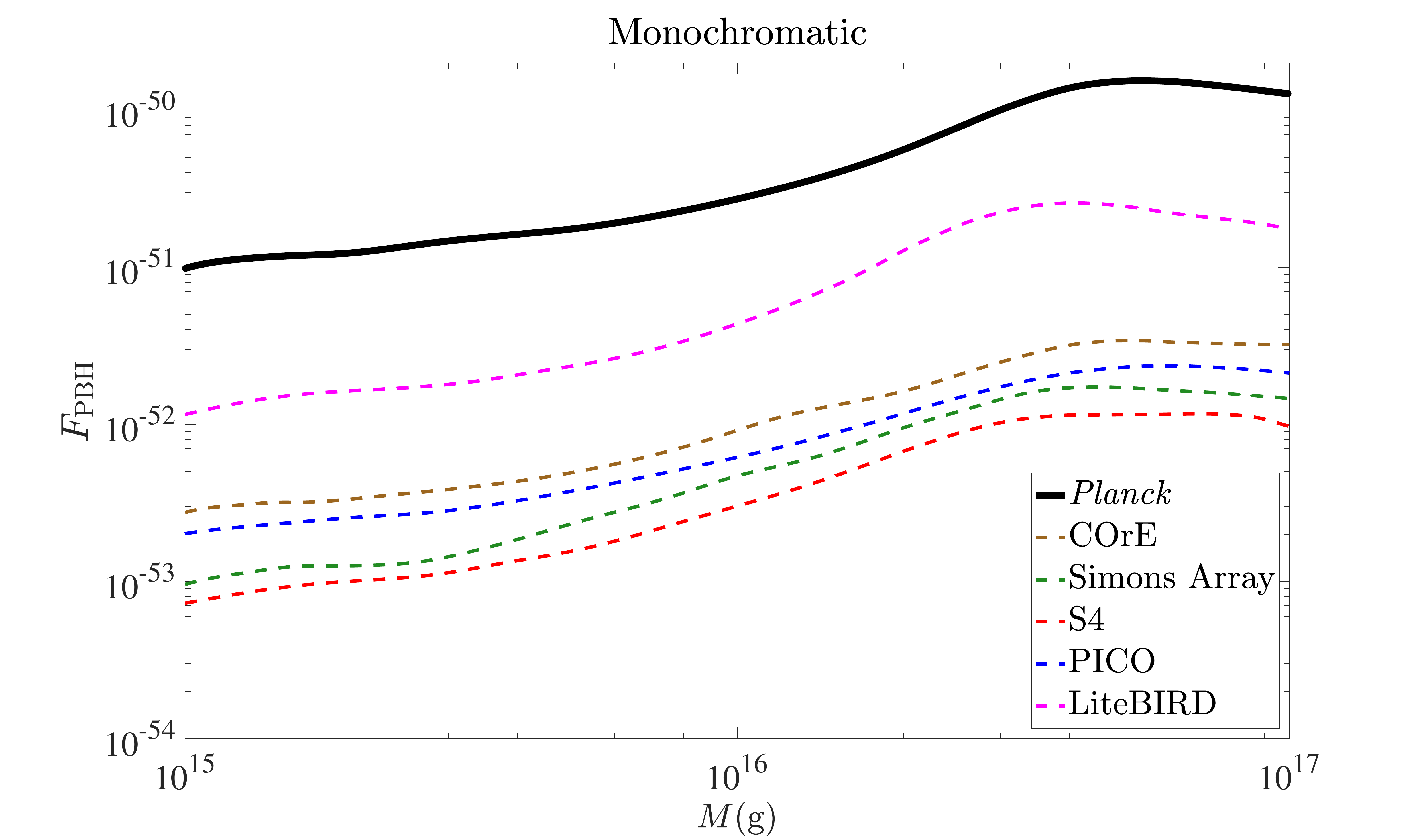}
\caption{Prospective 95\% C.L upper bounds on $F_{\rm{PBH}}$ defined in Eq.(\ref{Def_F}).
}
\label{F_Mono}
\end{figure}

\begin{figure*}[htp] 
\subfigbottomskip=-200pt
\subfigcapskip=-7pt
\subfigure{\includegraphics[width=8cm]{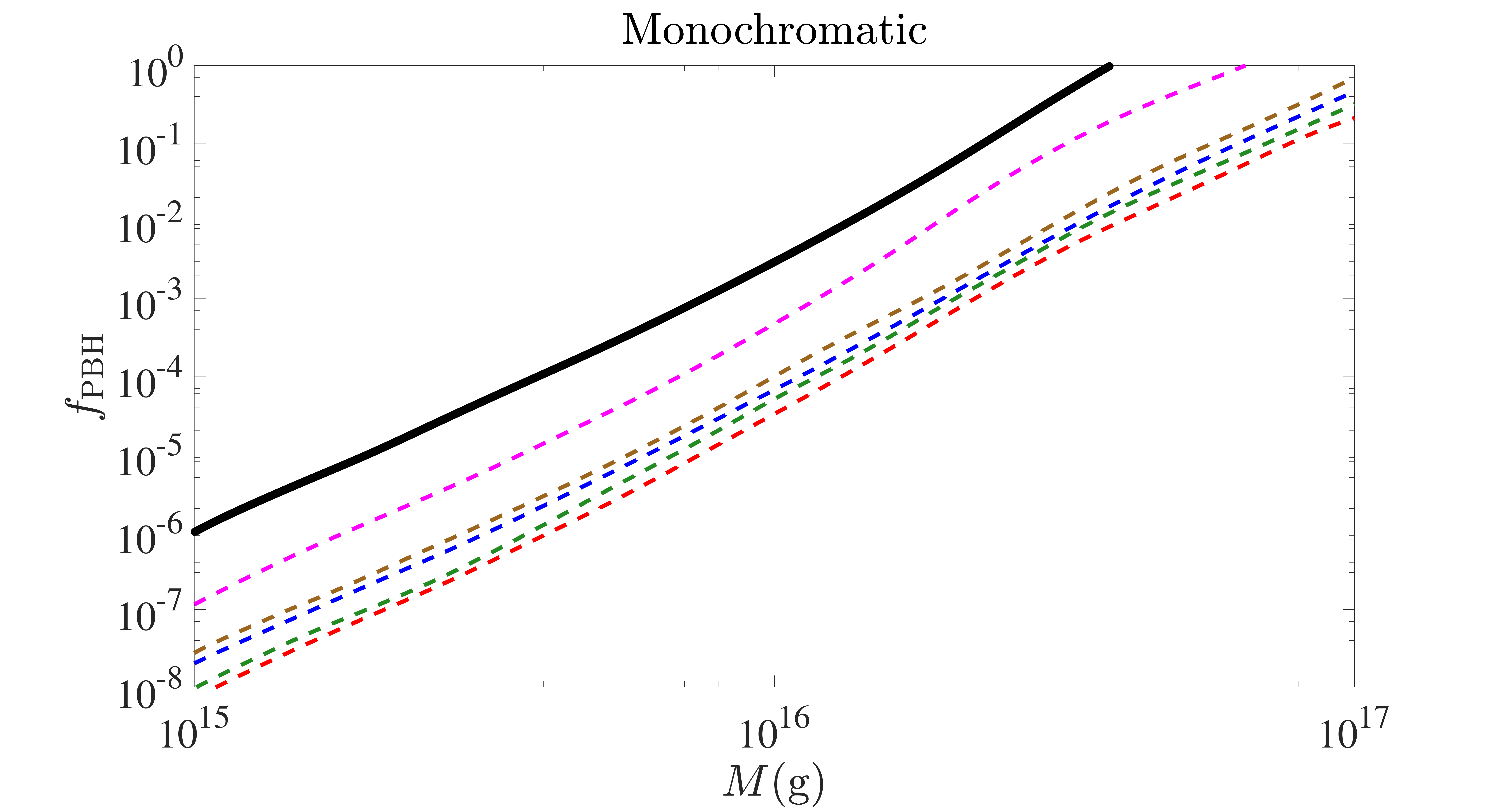}\includegraphics[width=8cm]{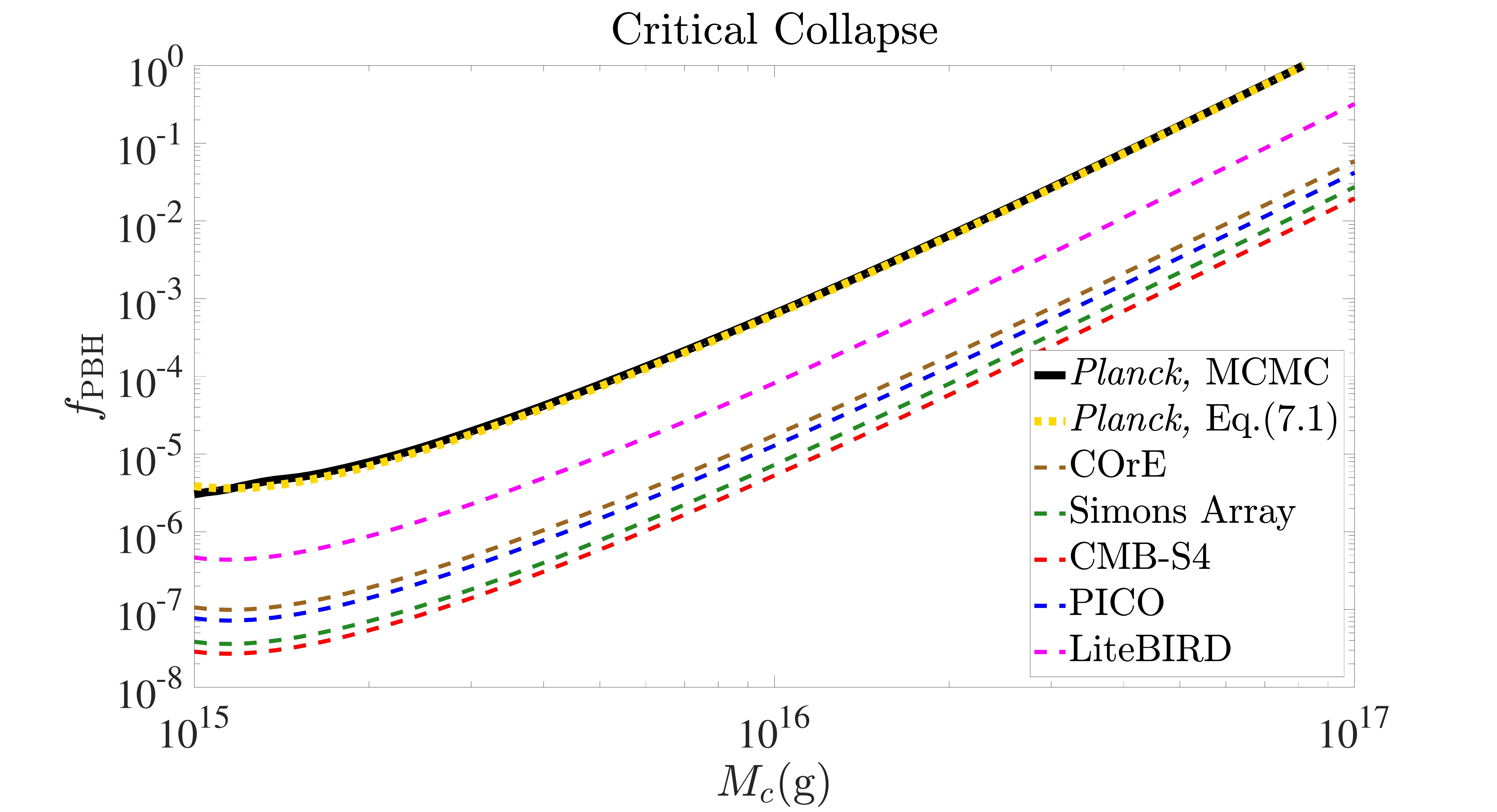}}
\subfigure{\includegraphics[width=8cm]{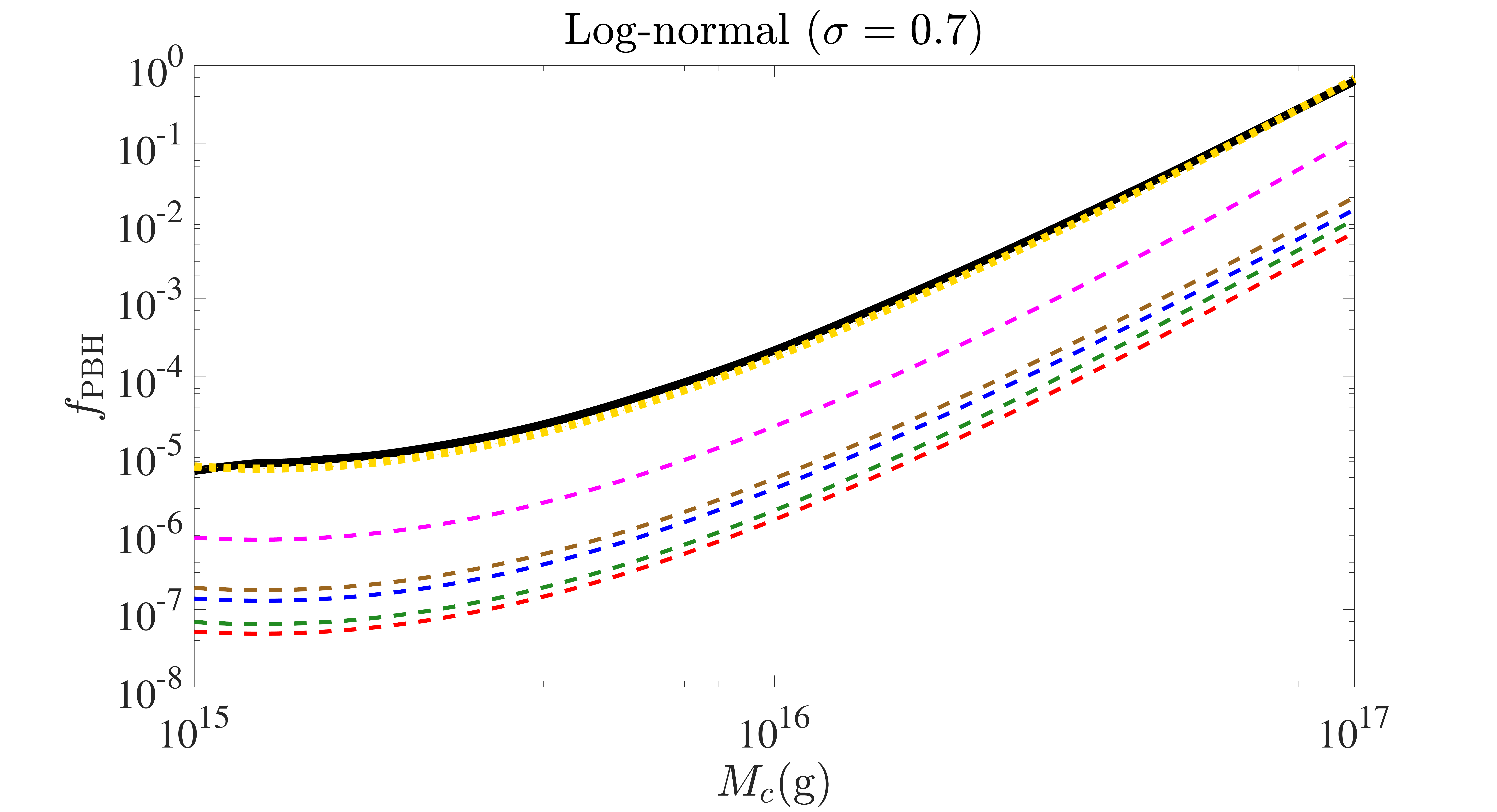}\includegraphics[width=8cm]{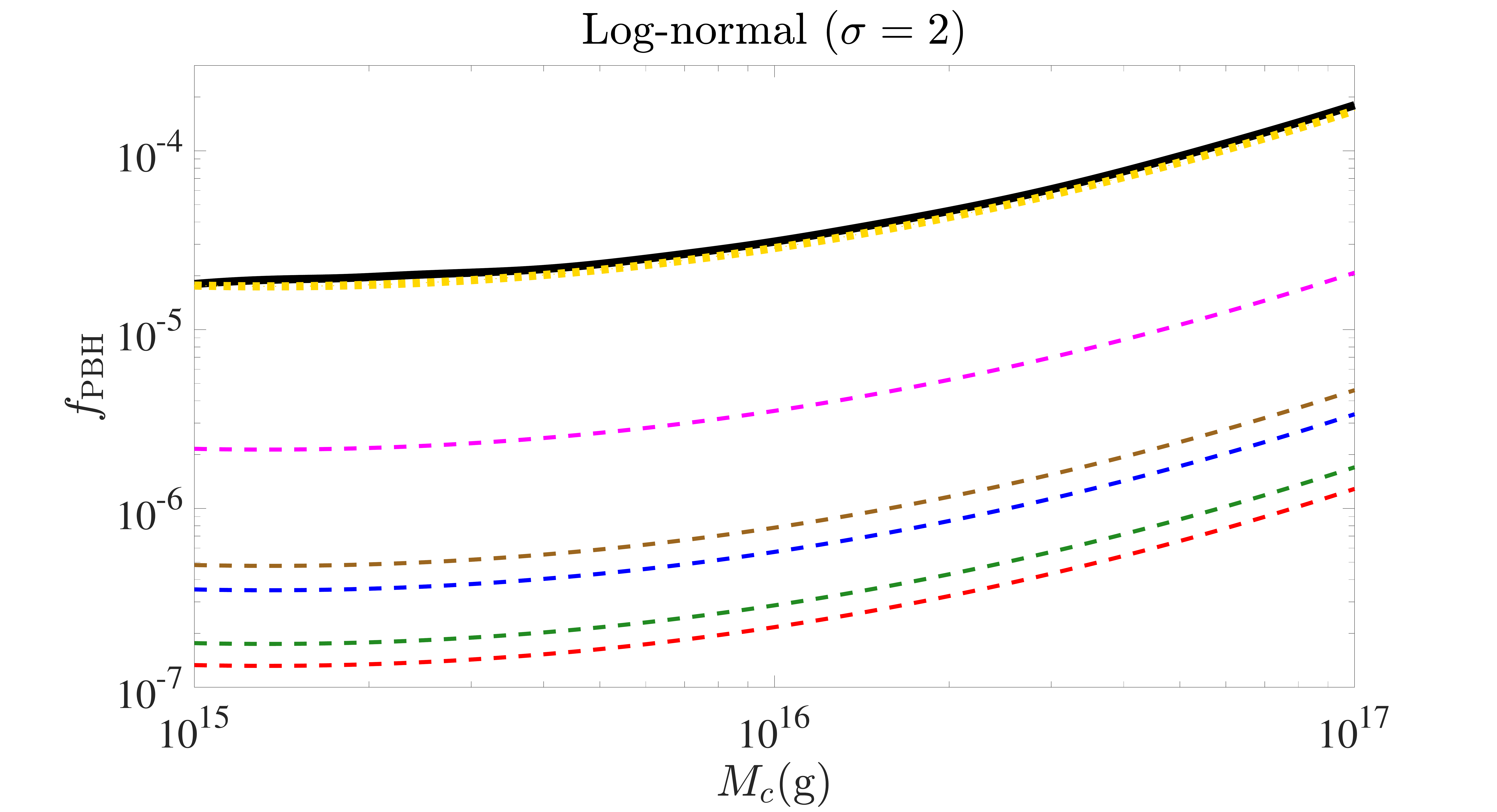}}
\subfigure{\includegraphics[width=8cm]{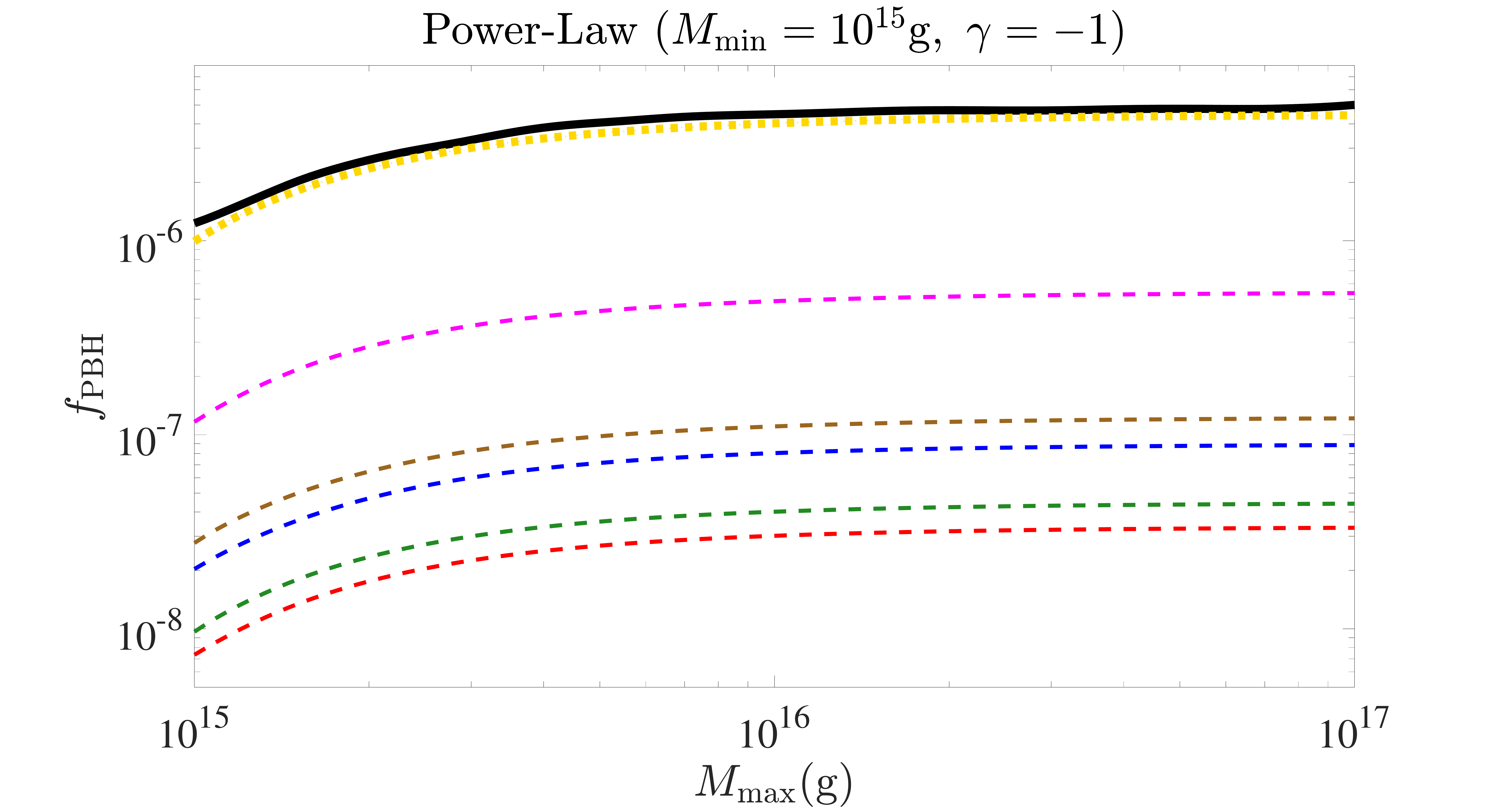}\includegraphics[width=8cm]{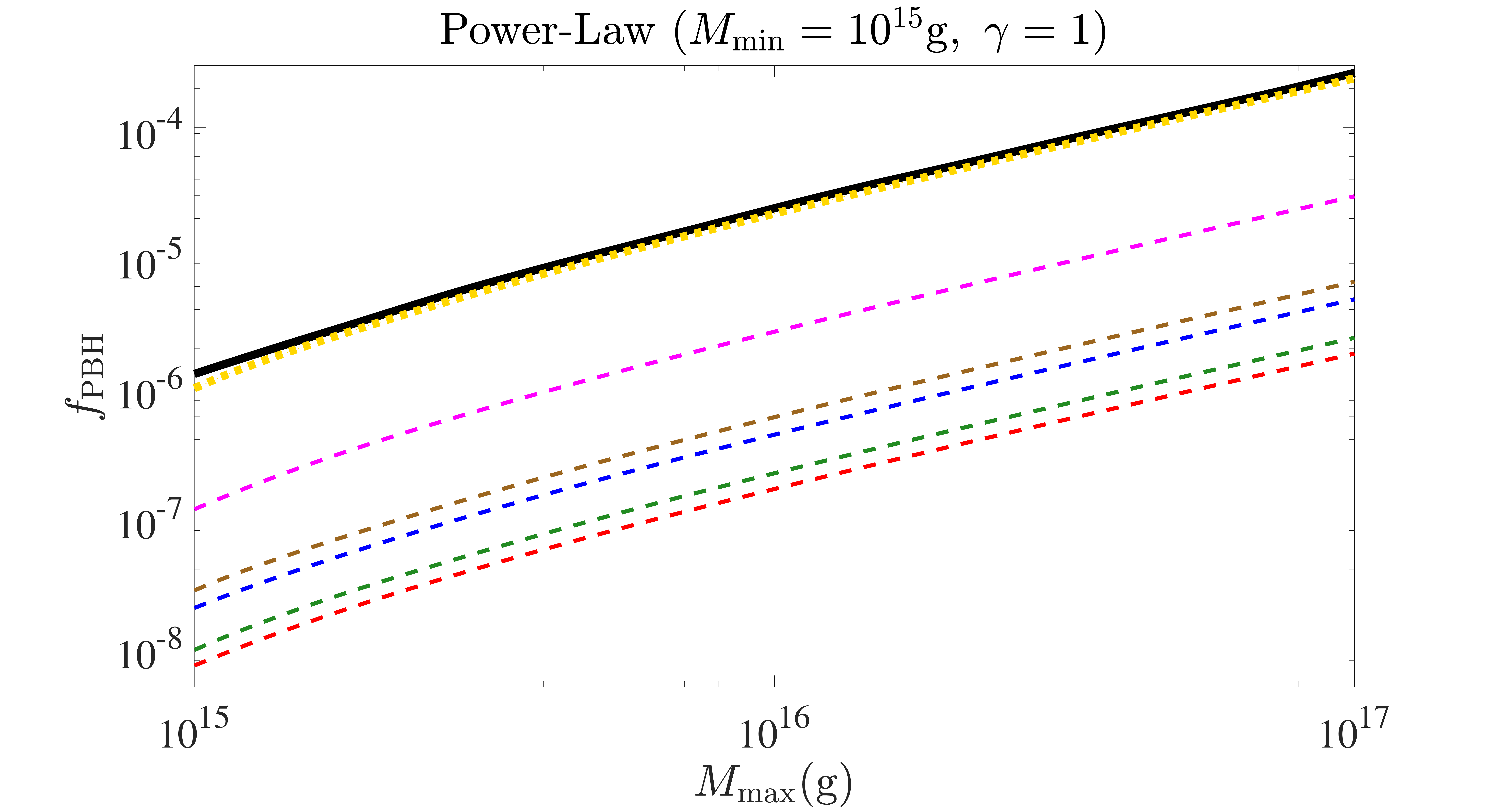}}
\subfigure{\includegraphics[width=8cm]{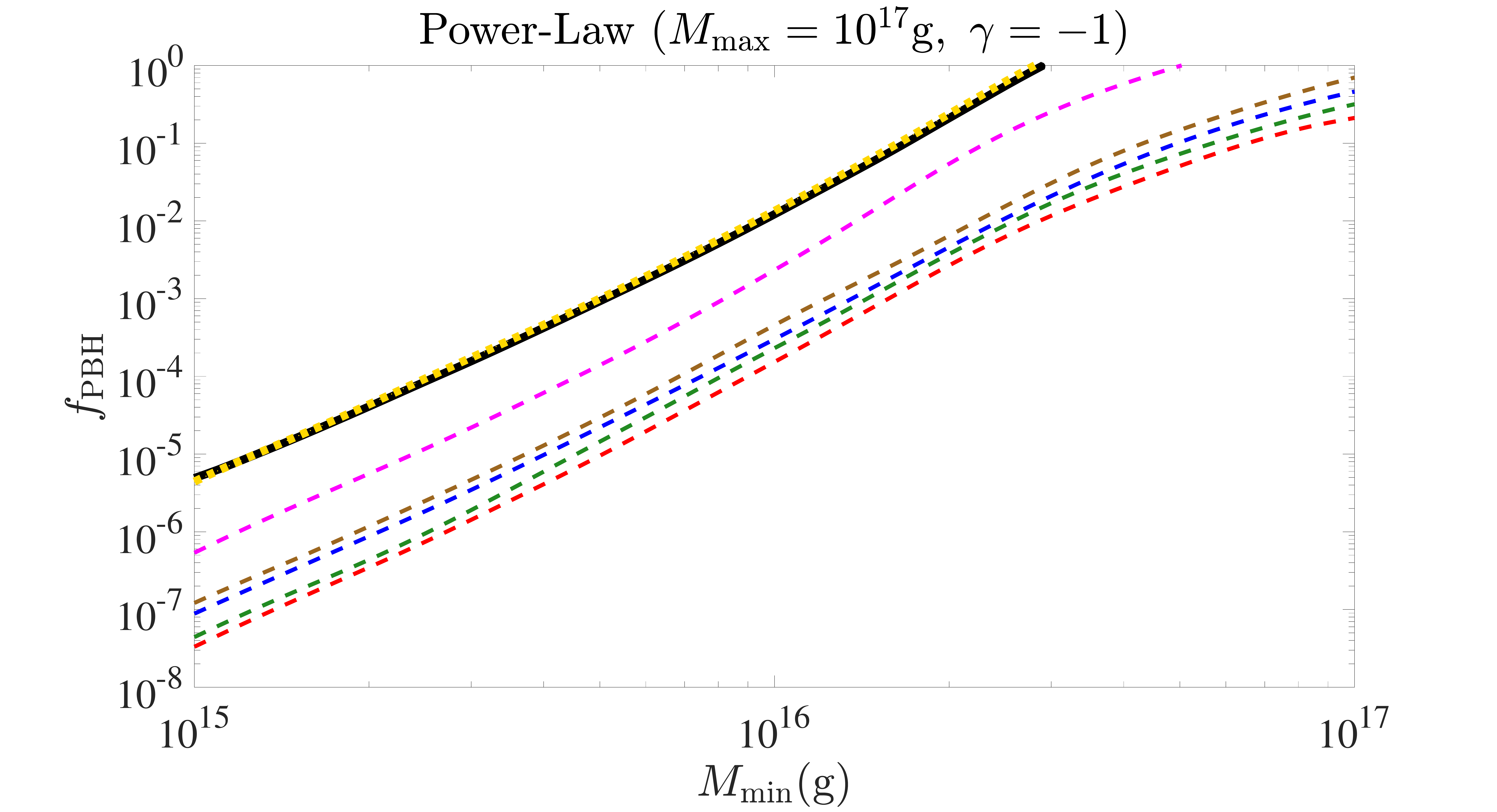}\includegraphics[width=8cm]{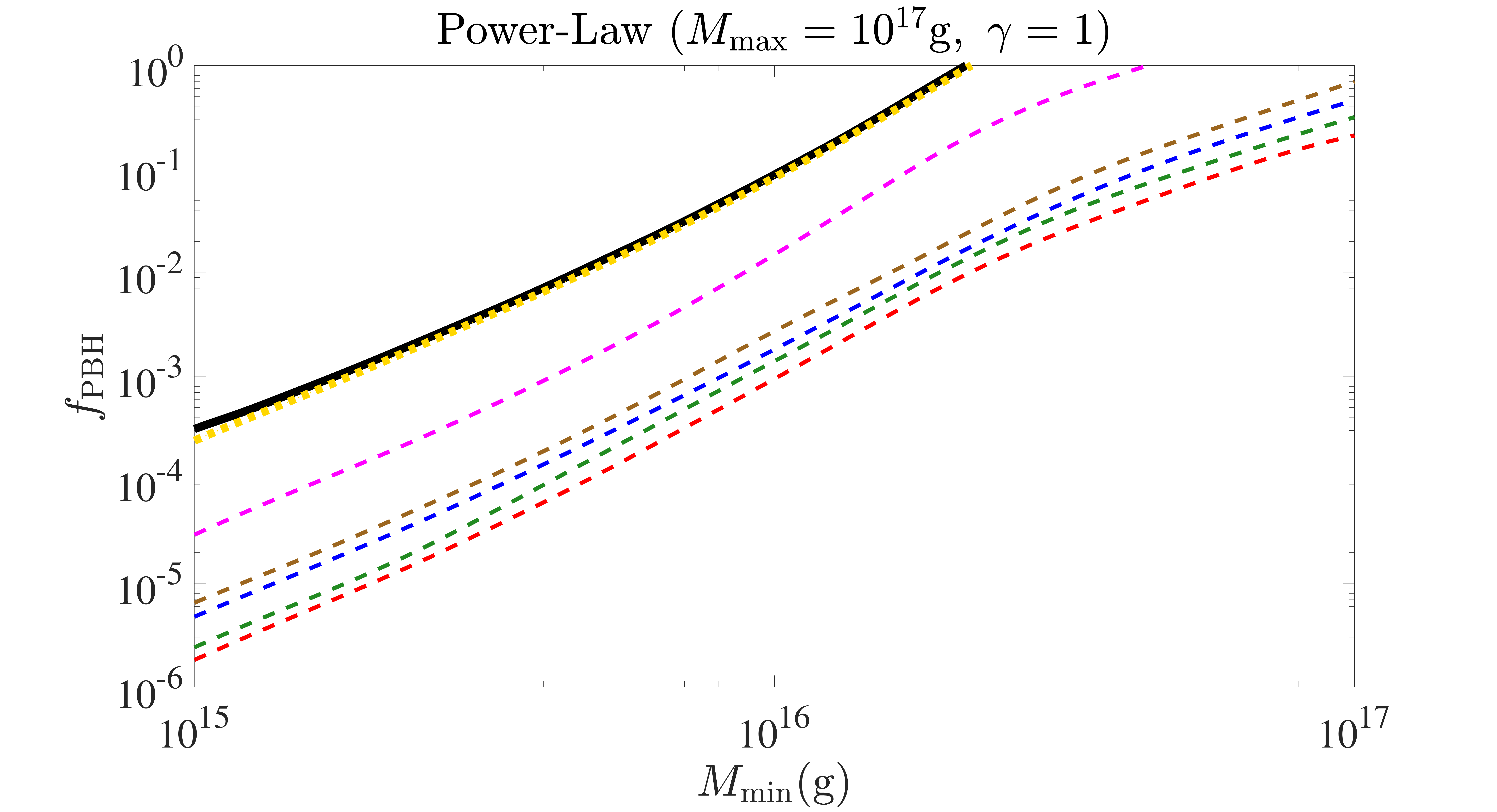}}
\subfigure{\includegraphics[width=8cm]{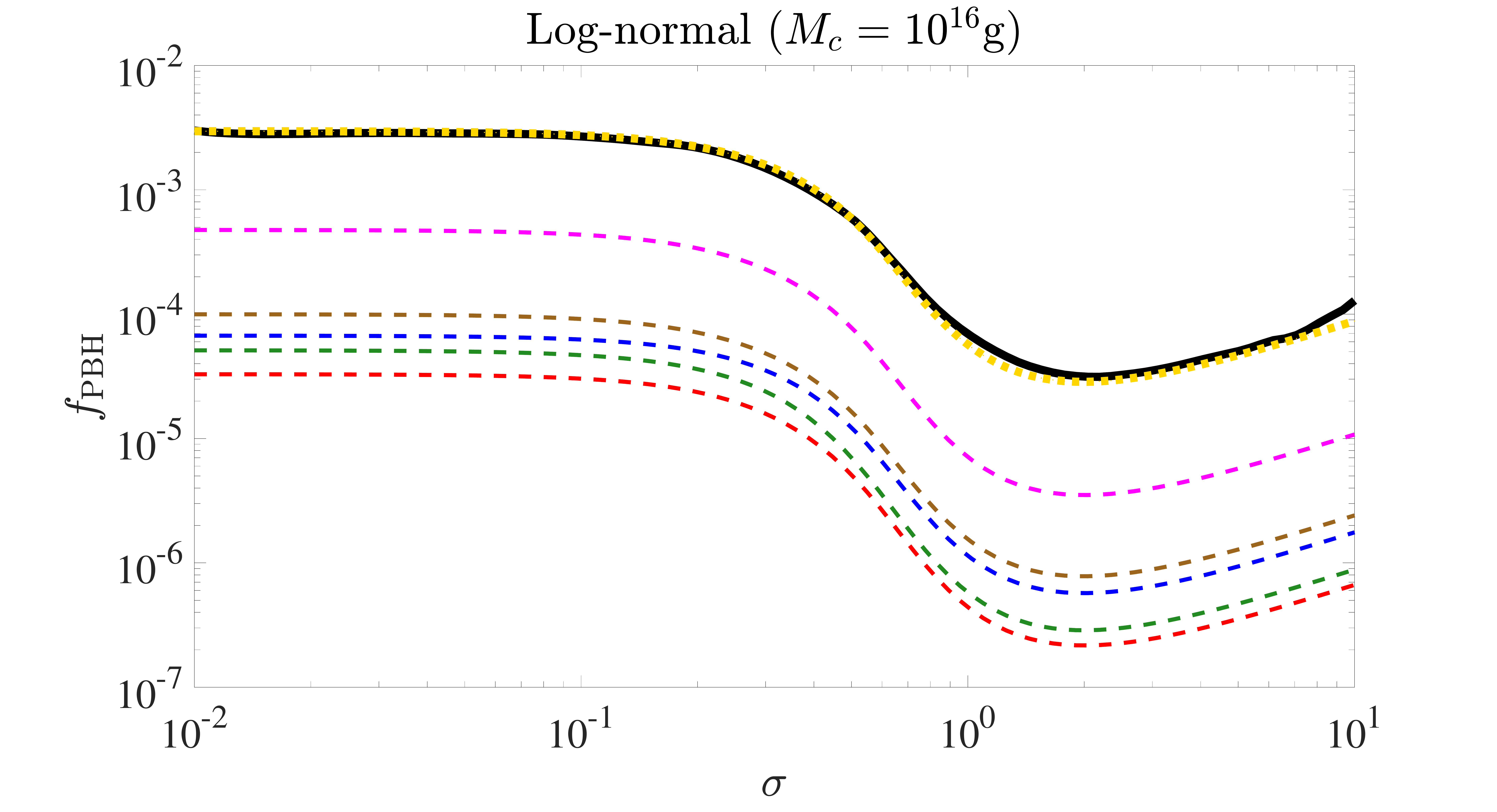}\includegraphics[width=8cm]{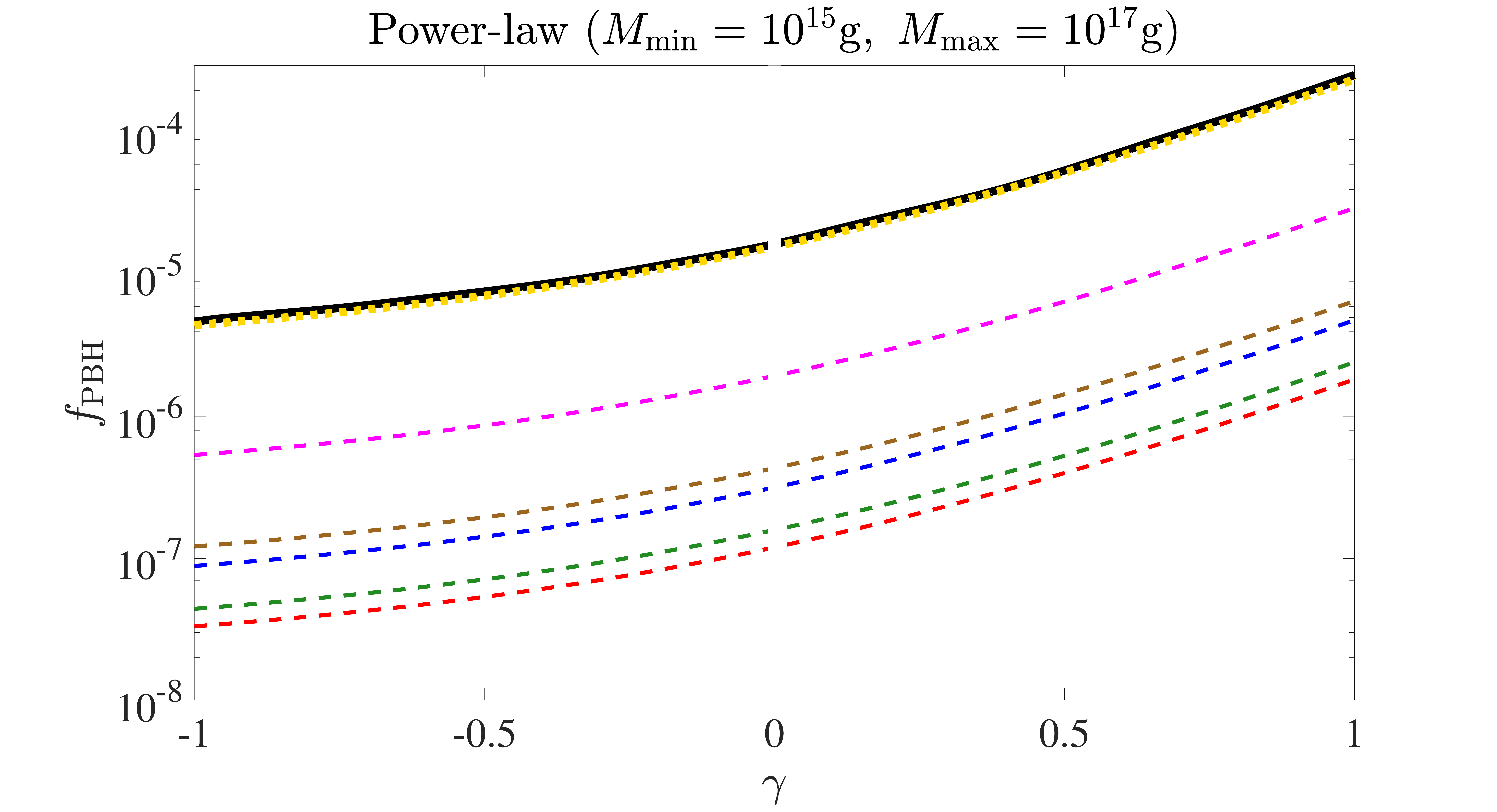}}
\caption{Prospective 95\% C.L upper bounds on $f_{\rm{PBH}}$, legend applies to all panels.
For extended distributions, {\it{Planck}} constraints derived from MCMC analysis and Eq.~(\ref{Extended_Constraints_Re_Interpret}) are shown in black solid and yellow dotted lines respectively.
}
\label{Result_1D}
\end{figure*}

We use modified HyRec~\cite{AliHaimoud:2010dx} and CAMB~\cite{Lewis:1999bs} codes to solve for recombination history and resulting CMB anisotropy spectrum. 
In addition to forecasted results for future experiments, we also show constraints set by $\it{Planck}$ 2018 data~\cite{Aghanim:2019ame}: (i) the plik-lite TTTEEE likelihood for $\ell > 29$, (ii) the low-$\ell$ TT and EE likelihoods for $\ell < 29$, (iii) lensing likelihood. 
All monochromatic constraints,
shown in Fig.\ref{F_Mono} and the top left panel in Fig.\ref{Result_1D}, 
are obtained by sampling our likelihoods using 
CosmoMC~\cite{Lewis:2002ah,Lewis:2013hha}.
The six $\rm{\Lambda CDM}$ parameters are varied in MCMC analysis and marginalized in our results.

\begin{figure*}[htp] 
\centering
\subfigbottomskip=-200pt
\subfigcapskip=-10pt
\subfigure[{Log-normal}]{\includegraphics[width=8cm]{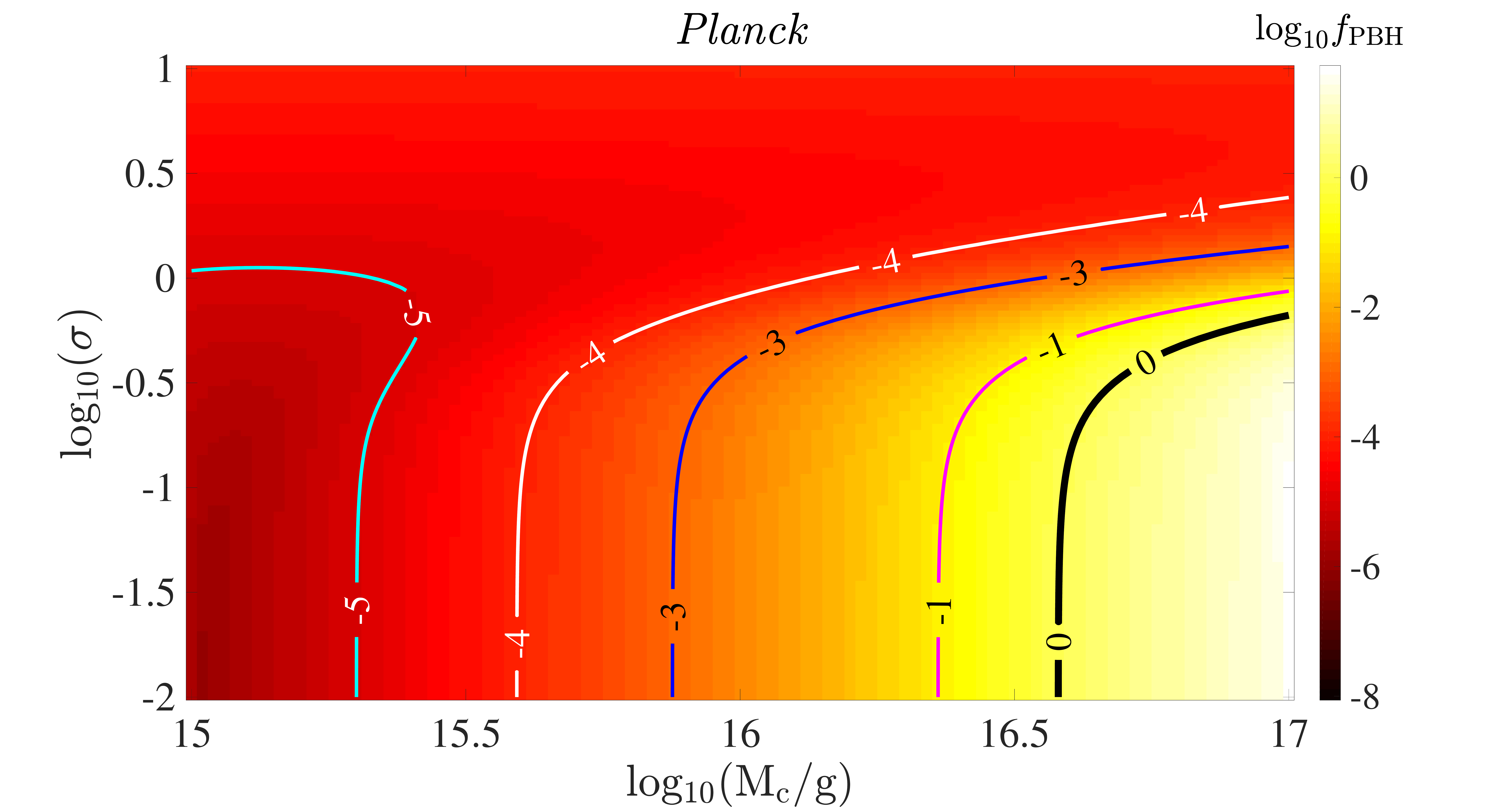}\includegraphics[width=8cm]{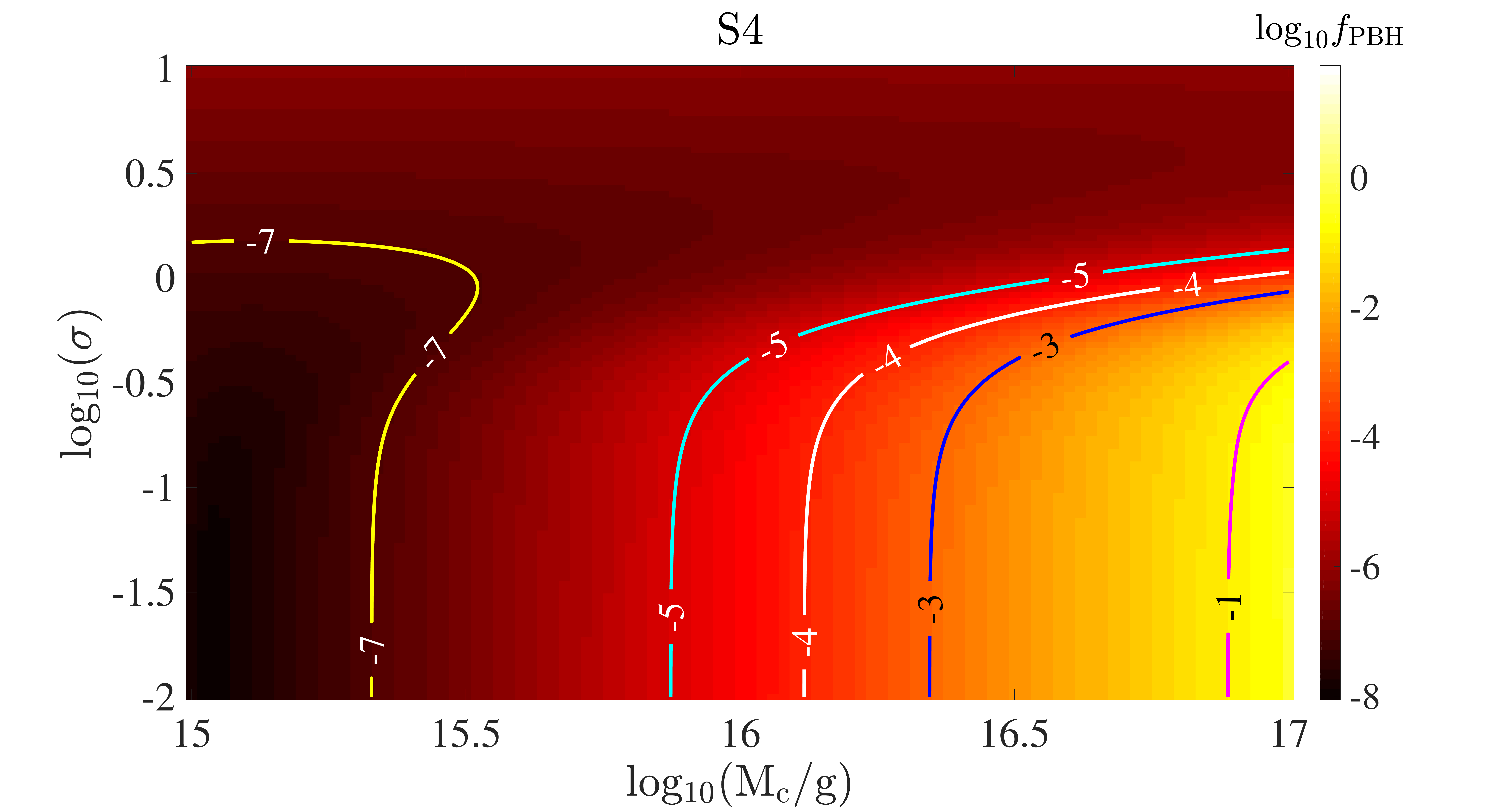}}
\subfigure[{Power-law, $M_{\rm{min}}=10^{15}{\rm{g}}$}]{\includegraphics[width=8cm]{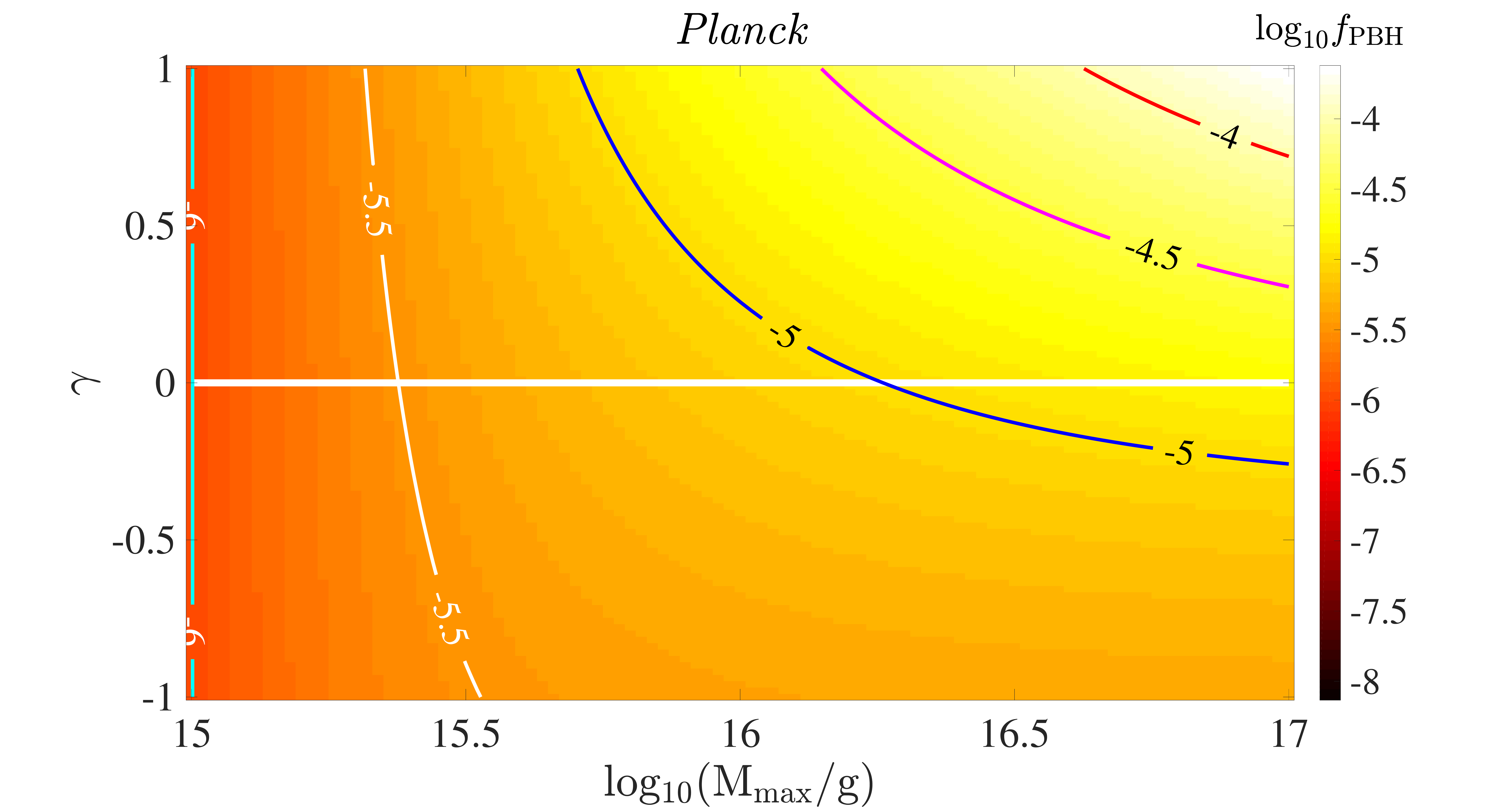}\includegraphics[width=8cm]{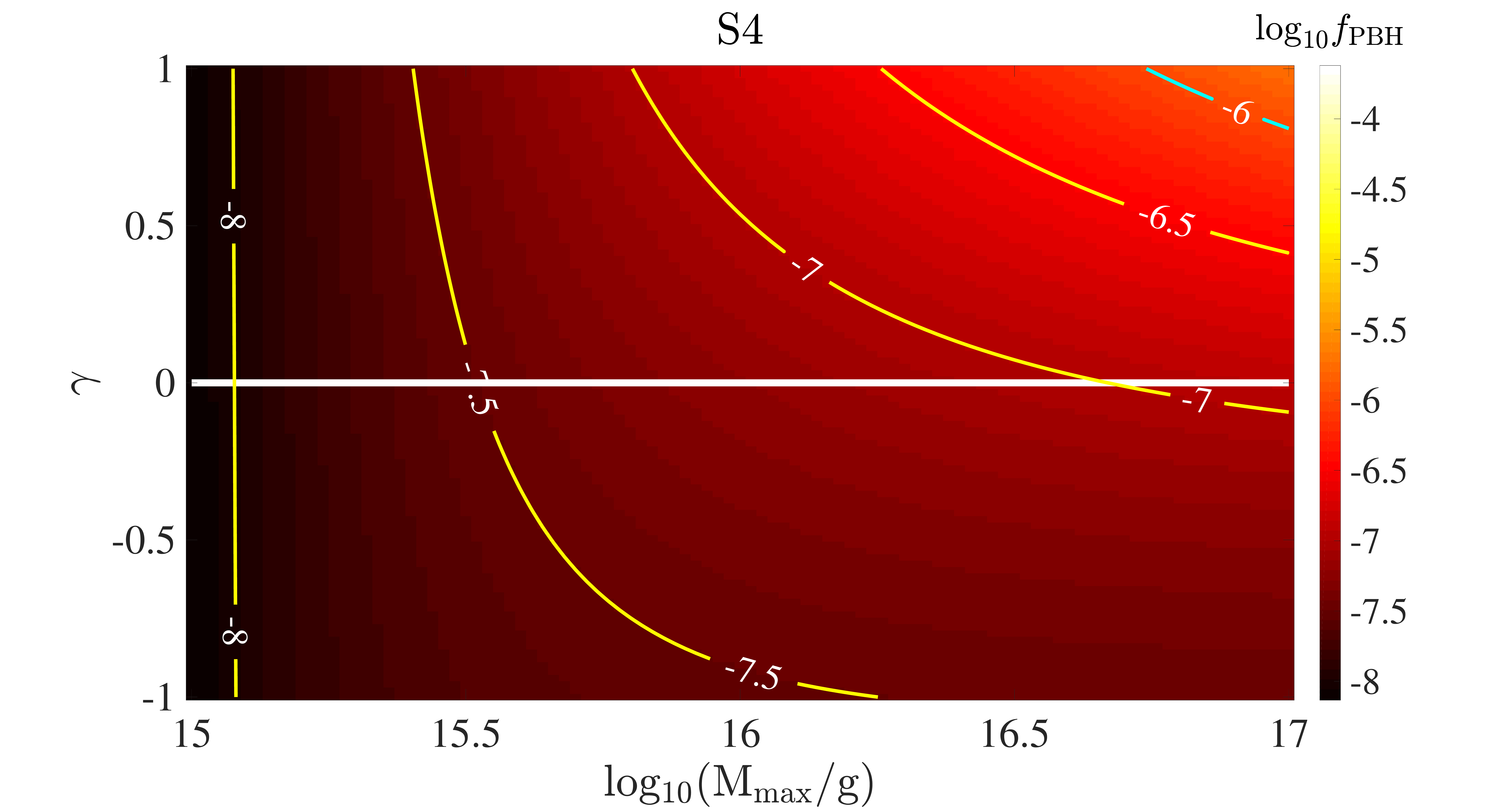}}
\subfigure[{Power-law, $M_{\rm{max}}=10^{17}{\rm{g}}$}]{\includegraphics[width=8cm]{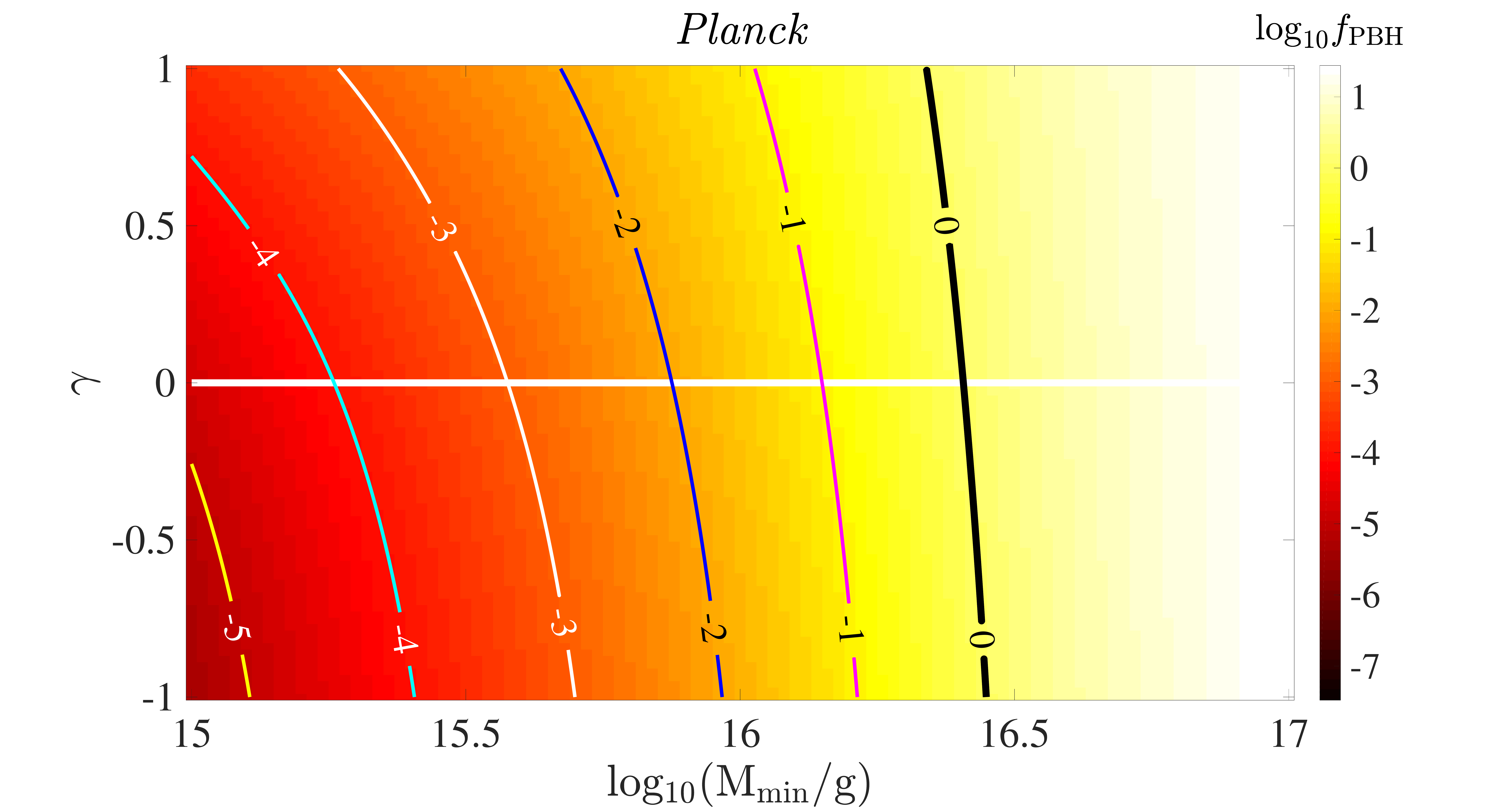}\includegraphics[width=8cm]{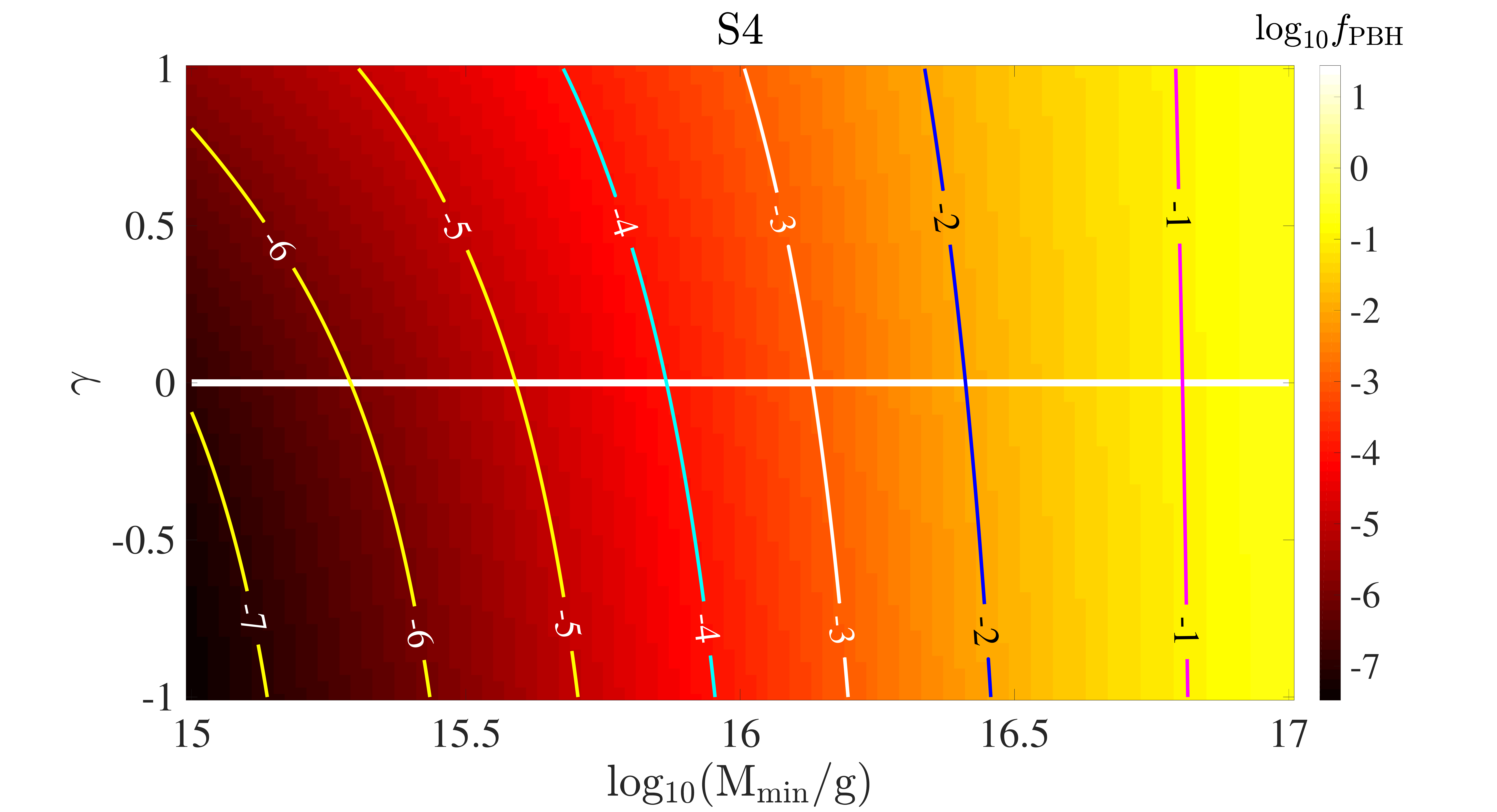}}
\subfigure[{Power-law, $\gamma=1$}]{\includegraphics[width=8cm]{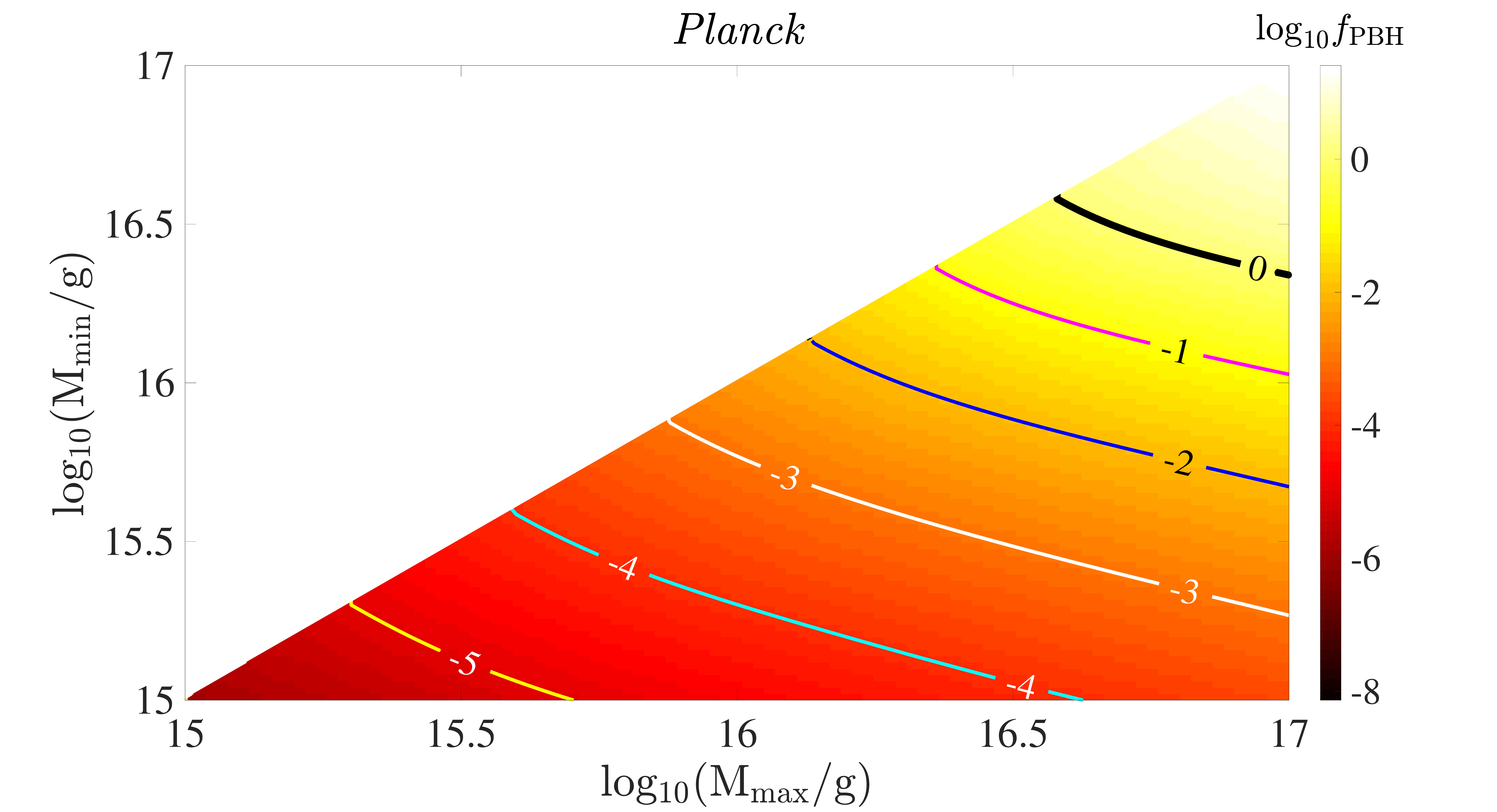}\includegraphics[width=8cm]{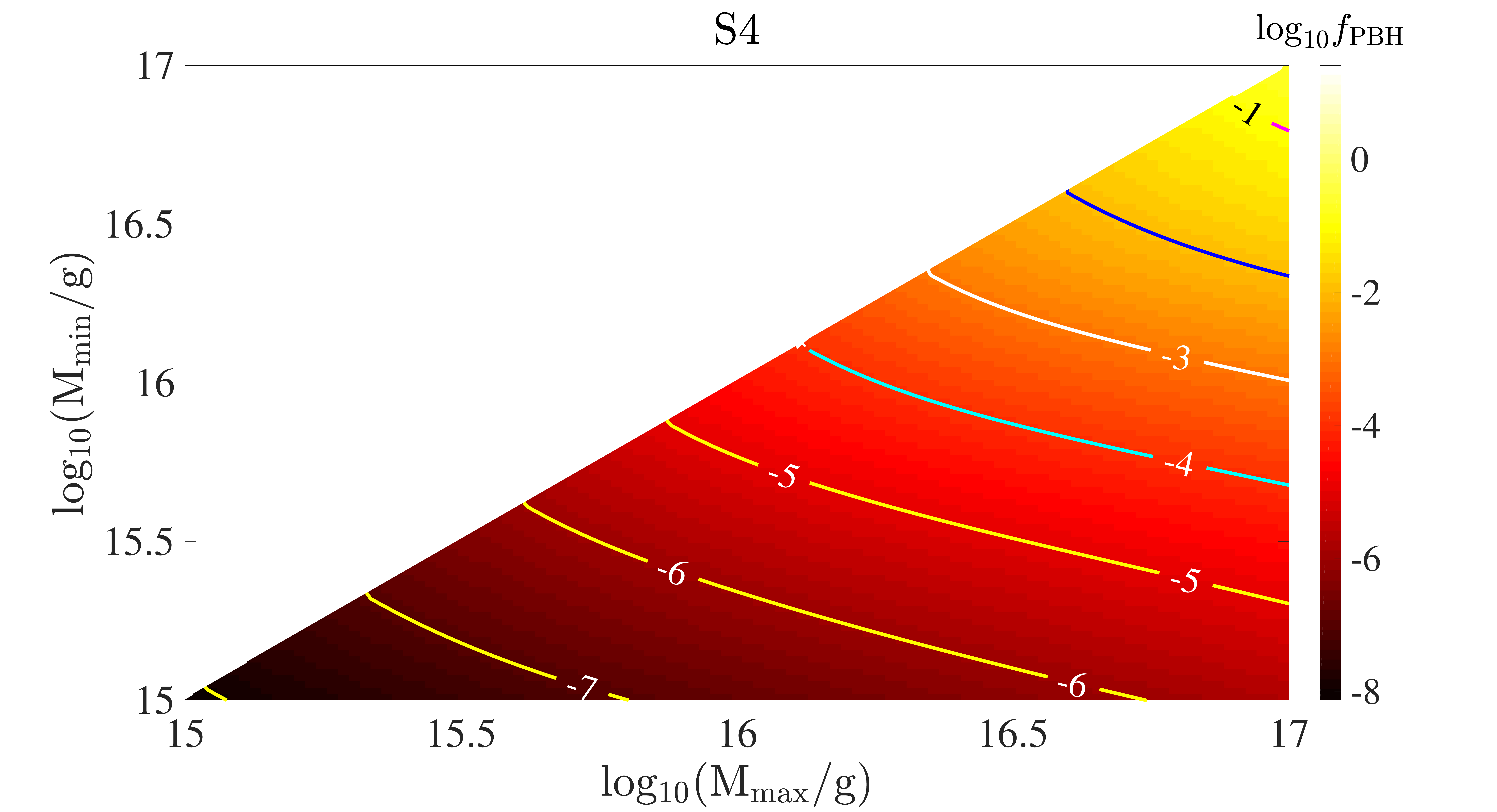}}
\subfigure[{Power-law, $\gamma=-1$}]{\includegraphics[width=8cm]{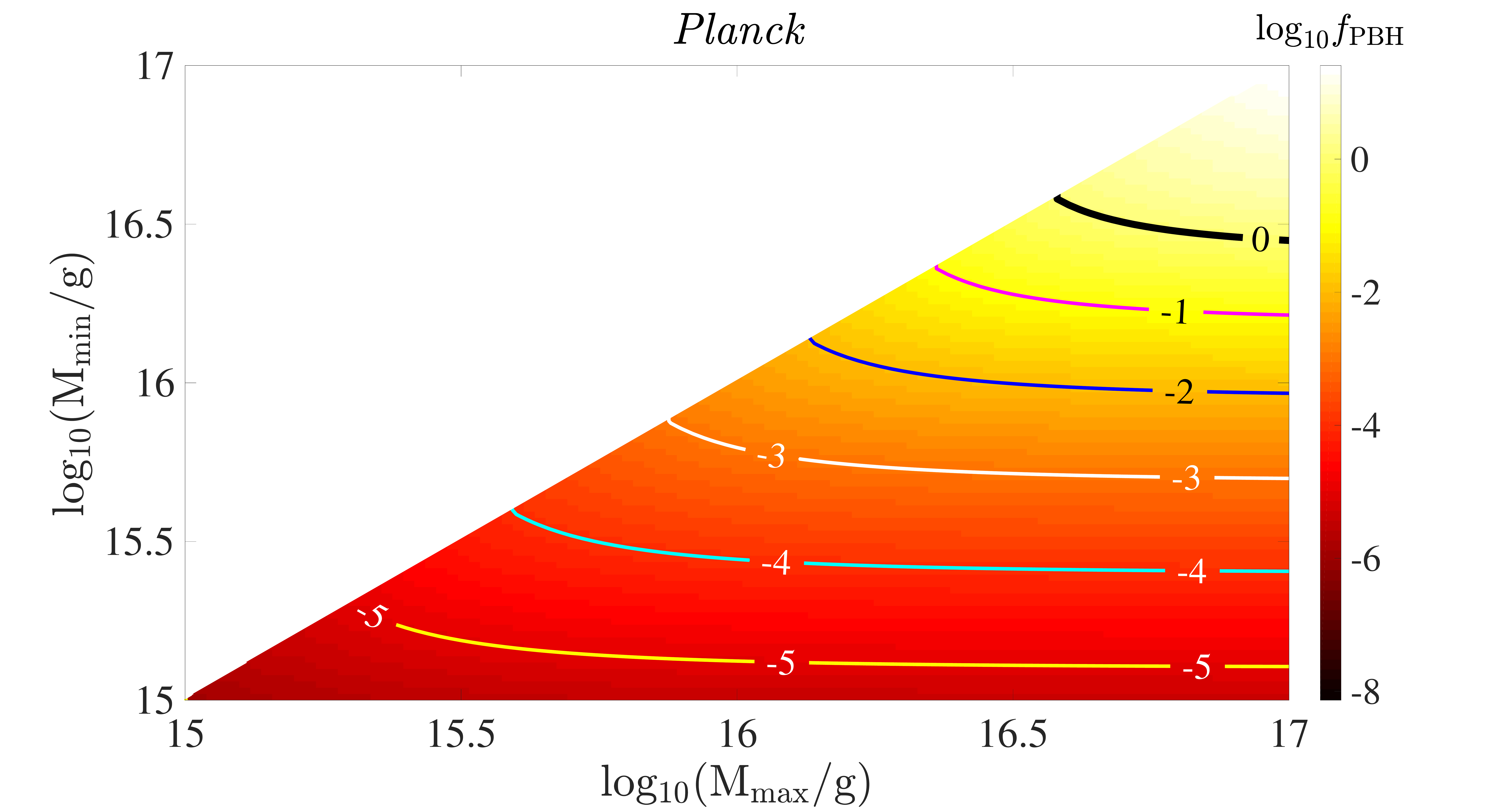}\includegraphics[width=8cm]{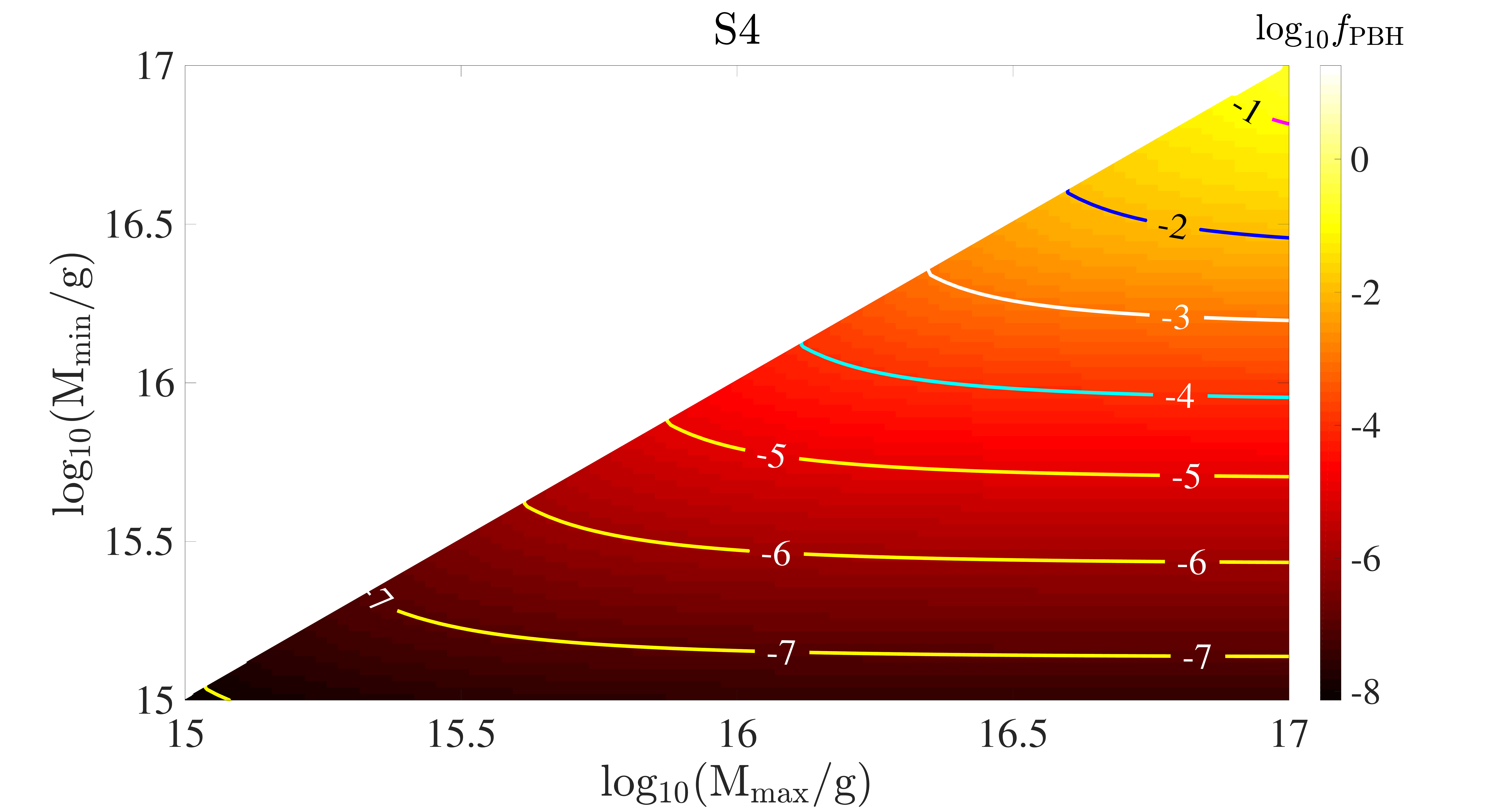}}
\caption{
{\it{Planck}} (left) and projected CMB-S4 (right) 95\% C.L upper bounds on ${\rm{log}}_{10}f_{\rm{PBH}}$.
Black contours indicates regions where PBHs can account for all of DM.
The white horizontal lines in panels b and c indicate $\gamma=0$, which is not considered in this work.
}
\label{Result_2D}
\end{figure*}

For computational convenience, here we calculate $f_{\rm{PBH}}$ upper bounds for extended distributions using \cite{Carr:2017jsz,Kuhnel:2017pwq}
\bea 
\label{Extended_Constraints_Re_Interpret}
f_{\rm{PBH}}
\le
1/
\int^{10^{17}{\rm{g}}}_{10^{15}{\rm{g}}}
 {\rm{d}} M
\frac
{\Psi(M)}
{f_{\rm{max}} (M)},
\eea
where $f_{\rm{max}} (M)$ is our MCMC bounds on $f_{\rm{PBH}}$ for the monochromatic PBHs with mass $M$.
This method was proven exact provided that $f_{\rm{max}} (M)$ were derived from one single observable\cite{Carr:2017jsz}.
For all our crosschecks with MCMC results for {\it{Planck}}, which has also been included in Fig.\ref{Result_1D} in black solid lines, we found that constraints derived using this equation agrees very well with that given by full MCMC simulation (at 10\% level).

As shown in Figs.\ref{Result_1D} and \ref{Result_2D},
for all PBH distributions we considered, significantly improved $f_{\rm{PBH}}$ constraints are expected from future CMB observations. 
For monochromatic and extended mass functions, there are still regions in parameter space allowed by current data where PBH can serve as the dominant ($f_{\rm{PBH}} \sim 1$) DM component. 
For example, monochromatic PBHs heavier than $3.8 \times 10^{16} {\rm{g}}$ can still account for all DM in the universe according to {\it{Planck}}.
Future missions are capable of testing all  these possibilities.
The Simons Array~\cite{Arnold:2014qym,Creminelli:2015oda}, which has already started taking data, along with proposed missions such as PICO~\cite{Hanany:2019wrm,Young:2018aby} and CMB-S4~\cite{Abazajian:2019eic,Abazajian:2016yjj}, can constrain monochromatic PBH abundance down to $F_{\rm{PBH}} \sim 10^{-53}$, improved by about two orders of magnitudes compared with ${\it{Planck}}$, and about one order of magnitude more stringent than the EBG bounds~\cite{Carr:2016drx,Carr:2009jm}.
We also find that our monochromatic {\it{Planck}} bound is in good agreement with that in ~\cite{Stocker:2018avm}.

In general, as lighter PBHs are more radiant, $f_{\rm{PBH}}$ constraints weakens as $M$ increases. 
By choosing $F_{\rm{PBH}}$ to parameterize monochromatic PBH injection instead of $f_{\rm{PBH}}$, we factored out the dependence on the $M^3$ term and massive particle emission fraction ($\omega^{\rm{e}^{\pm}} (M)$) in Eq.(\ref{Injection_Rate}). 
As a result, the dependence of $F_{\rm{PBH}}$ constraints on PBH mass in Fig.\ref{F_Mono} 
 is determined mostly by the way in which the deposition efficiency relates to $M$. 
A visible peak on $F_{\rm{PBH}}$ can be seen at around $4 \times 10^{16} \rm{g}$, which roughly corresponds to regions where the deposition efficiency is at the lowest.

For monochromatic PBHs, the EGB upper bound on $f_{\rm{PBH}}$ can be parameterized by\cite{Carr:2016drx},
\bea \label{EGB_Bound}
f^{\rm{EGB}}=2 \times 10^{-8}
\left(\frac{M}{M_{\star}}\right)^{3+\epsilon}
\ (M_{\star} < M < 10^{18}{\rm{g}}),
\eea
where $M_{\star}=5 \times 10^{14} {\rm{g}}$ is the mass of a PBH whose lifetime equals the age of the Universe. 
$\epsilon$ parameterizes the spectral index of extragalactic intensity and has values between $0.1$ and $0.4$.
Similarly we found that our CMB bound can be well approximated by

\bea 
f^{\rm{CMB}}=
\frac{4.32}{K}
\times 10^{-8}
\left(\frac{M}{M_{\star}}\right)^{3+\epsilon '}
\ (10^{15}{\rm{g}} < M < 10^{17}{\rm{g}}),
\label{CMB_Bound}
\eea
where $\epsilon ' =0.84$, $K$ is an experiment specific scaling factor that describes how stringent each mission constrains $f_{\rm{PBH}}$. $K$ is normalized to unity for {\it{Planck}} and is generally larger for forecasted missions. The prospective value of this scaling factor for future experiments are listed in Table.\ref{K_Factor}. The experiments' specifications are listed in Appendix~\ref{Exp_Specs}.  

\begin{table}[ht]
\begin{center}
\begin{tabular}{c|c}
\hline
        Experiment 	      &  Scaling Factor\\
        \hline
	{\it{Planck}}                                &1\\ 
	COrE                                         &37\\
	CMB-S4                                    &113\\ 
	PICO                                         &53\\
	LiteBIRD                                    &7\\
	Simons Array                             &80\\
\hline
\end{tabular}
\caption{
The best-fit scaling factor $K$ defined in Eq.~(\ref{CMB_Bound}), for experimental specifics and $\ell$ ranges listed in Tab~\ref{Tab_ExpSpecs}.
}
\label{K_Factor}
\end{center}
\end{table}

As expected, we recover the monochromatic constraints when $\sigma \to 0$ in log-normal model or $M_{\rm{min}} \to M_{\rm{max}}$ in power-law model.
For the log-normal distribution, $\sigma=0.2$ can already give a good approximation for monochromatic $f_{\rm{PBH}}$ bound, with an average deviation of about 27\%.
We find that so long as the majority ($\ge 68\% $) of PBHs density are covered in our targeted $[10^{15},10^{17}] {\rm{g}}$ mass range, increasing the distribution width $\sigma$ in log-normal model tightens $f_{\rm{PBH}}$ constraints,
which is in agreement with the claims in Refs.\cite{Carr:2017jsz}.
PBHs with $\sigma$ larger than 0.66 are completely ruled out as the dominant DM component by {\it{Planck}} data.
However for regions where an appreciable amount of PBH density are distributed outside the $[10^{15},10^{17}]$ g window,
$f_{\rm{PBH}}$ bound relaxes as we increase $\sigma$.
For power-law distribution, we find that higher $\gamma$, which corresponds to wider distributions, allows for higher $f_{\rm{PBH}}$.
For $\gamma=1$, the power-law mass function automatically reduces to an uniform distribution in $[M_{\rm{min}}, M_{\rm{max}}]$ with $\Psi={\rm{Const}}$, 
results for which are also presented in Figs.\ref{Result_1D} and \ref{Result_2D}.

Note that in case of null-signal, whether extending the PBH distribution tighten the $f_{\rm{PBH}}$ upper bounds depends on whether the extension leads to more PBH radiation.
For the log-normal model, if the parameters in the mass function still ensure that a significant fraction of PBH density is contained in the mass range that can be probed through Hawking radiation, then increasing the distribution width will continuously include smaller, more radiant PBHs, which in turn tightens $f_{\rm{PBH}}$ bounds.
Otherwise if the majority of PBHs were distributed outside the probed mass window, then the extension will include less PBH radiation and the constraints on $f_{\rm{PBH}}$ will be relaxed,
making it possible to evade monochromatic bounds by merely extending the distribution.

\section{Summary} \label{Discussion}
Hawking radiation from primordial black holes during the cosmic dark ages can ionize and heat the intergalactic medium. 
This effect increases the scattering between CMB photons and free electrons, thereby changing the CMB anisotropy spectrum.  
In this paper we forecast the sensitivity of several future CMB experiments in constraining PBHs in $10^{15} \sim 10^{17} {\rm{g}}$ mass range. 
We find that future experiments can significantly improve current CMB bounds and 
rule out PBHs as the dominant DM component for all mass functions considered.
For the conventional monochromatic distribution, LiteBIRD constrains $f_{\rm{PBH}}/M^3$ to orders of $10^{-52}{\rm{g^{-3}}}$,
bounds from COrE, PICO, Simons Array and CMB-S4 can extend down to $f_{\rm{PBH}}/M^3 \sim 10^{-53}{\rm{g^{-3}}}$, improved by about two orders of magnitudes compared to {\it{Planck}}.
We also considered three extended distribution models: log-normal, power-law and critical collapse.
Provided that at least $68\%$ of PBH density are still contained in $[10^{15},10^{17} ]$g, 
for log-normal models with large distribution width $\sigma$ and the critical collapse distribution,
$f_{\rm{PBH}}$ bound tightens relative to the monochromatic results. 
Most of our extended constraints are obtained by re-interpreting the monochromatic bounds, and we have cross-checked that these results are consistent with that derived from full MCMC analysis.

\medskip
{\bf Acknowledgements}
J.C. is supported by the Sino US Cooperation Project of Ministry of Science and Technology (No. 2016YFE0104700).
Y.G. is partially supported by the Institute of High Energy Physics, Chinese Academy of Sciences, under the grant no. Y7515560U1.
YZM acknowledges National Research Foundation of South Africa with grant no. NRF-120385, NRF-120378, NRF-109577, and National Science Foundation of China with grant no. NSFC-11828301.

\newpage
\begin{appendix}
\section{Experimental Specifications} \label{Exp_Specs}

\begin{table*}[h]
\begin{center}
\begin{tabular}{lcccccc}
\toprule
\hline
${\rm{Experiment}}$  & $\ \ \ \ \ f_{\rm{sky}}$     & $\ \ \ \ \ell_{\rm{min}}$     & $\ \ \ \ \ \ell_{\rm{max}}$       &$\ \ \nu$                      & $\delta P$                      &$\theta_{\rm{FWHM}}$ \\
                                  &                                      &                                         &                                             & $\ \ {\rm{(GHz)}}$         & $(\mu{\rm{K}}$-arcmin)  &$(\rm{arcmin})$ \\
\hline

\multicolumn{1}{l}{\multirow{6}{*}{COrE ~\cite{DiValentino:2016foa,Delabrouille:2017rct}}} 	&\multicolumn{1}{l}{\multirow{6}{*}{\ \ \ \ \ \ 0.7}}      &\multicolumn{1}{l}{\multirow{6}{*}{\ \ \ \ \ 2}}         &\multicolumn{1}{l}{\multirow{6}{*}{\ \ \ \ \ 3000}}    & 90     & 7.3   & 12.1\\
\multicolumn{1}{l}{}                                                                                                                 & \multicolumn{1}{l}{}                                               & \multicolumn{1}{l}{}                                              & \multicolumn{1}{l}{}                                            & 100   & 7.1   & 10.9\\
\multicolumn{1}{l}{}                                                                                                                 & \multicolumn{1}{l}{}                                               & \multicolumn{1}{l}{}                                              & \multicolumn{1}{l}{}                                            & 115   & 7.0   & 9.6\\
\multicolumn{1}{l}{}                                                                                                                 & \multicolumn{1}{l}{}                                               & \multicolumn{1}{l}{}                                              & \multicolumn{1}{l}{}                                            & 130   & 5.5   & 8.5\\
\multicolumn{1}{l}{}                                                                                                                 & \multicolumn{1}{l}{}                                               & \multicolumn{1}{l}{}                                              & \multicolumn{1}{l}{}                                            & 145   & 5.1   & 7.7\\
\multicolumn{1}{l}{}                                                                                                                 & \multicolumn{1}{l}{}                                               & \multicolumn{1}{l}{}                                              & \multicolumn{1}{l}{}                                            & 160   & 5.2   & 7.0\\
\hline

\multicolumn{1}{l}{\multirow{2}{*}{CMB-S4 \cite{Abazajian:2019eic,Abazajian:2016yjj}}} 	&\multicolumn{1}{l}{\multirow{2}{*}{\ \ \ \ \ \ 0.62}}   &\multicolumn{1}{l}{\multirow{2}{*}{\ \ \ \ \ 30}}         &\multicolumn{1}{l}{\multirow{2}{*}{\ \ \ \ \ 3000}}    & 95     & 2.9   & 2.2\\
\multicolumn{1}{l}{}                                                                                                                 & \multicolumn{1}{l}{}                                               & \multicolumn{1}{l}{}                                              & \multicolumn{1}{l}{}                                              & 145   & 2.8   & 1.4\\
\hline

\multicolumn{1}{l}{\multirow{4}{*}{PICO ~\cite{Hanany:2019wrm,Hanany:2019lle,Young:2018aby}}} 	        &\multicolumn{1}{l}{\multirow{4}{*}{\ \ \ \ \ \ 0.7}}      &\multicolumn{1}{l}{\multirow{4}{*}{\ \ \ \ \ 2}}         &\multicolumn{1}{l}{\multirow{4}{*}{\ \ \ \ \ 4000}}    & 90     & 2.1   & 9.5\\
\multicolumn{1}{l}{}                                                                                                                 & \multicolumn{1}{l}{}                                               & \multicolumn{1}{l}{}                                              & \multicolumn{1}{l}{}                                            & 108   & 1.7   & 7.9\\
\multicolumn{1}{l}{}                                                                                                                 & \multicolumn{1}{l}{}                                               & \multicolumn{1}{l}{}                                              & \multicolumn{1}{l}{}                                            & 129   & 1.5   & 7.4\\
\multicolumn{1}{l}{}                                                                                                                 & \multicolumn{1}{l}{}                                               & \multicolumn{1}{l}{}                                              & \multicolumn{1}{l}{}                                            & 155   & 1.3   & 6.2\\
\hline

\multicolumn{1}{l}{\multirow{4}{*}{LiteBIRD \cite{LiteBIRD}}}                	                         &\multicolumn{1}{l}{\multirow{4}{*}{\ \ \ \ \ \ 0.7}}      &\multicolumn{1}{l}{\multirow{4}{*}{\ \ \ \ \ 2}}         &\multicolumn{1}{l}{\multirow{4}{*}{\ \ \ \ \ 200}}     & 89     & 11.7   & 35\\
\multicolumn{1}{l}{}                                                                                                                 & \multicolumn{1}{l}{}                                               & \multicolumn{1}{l}{}                                              & \multicolumn{1}{l}{}                                            & 100   & 9.2      & 29\\
\multicolumn{1}{l}{}                                                                                                                 & \multicolumn{1}{l}{}                                               & \multicolumn{1}{l}{}                                              & \multicolumn{1}{l}{}                                            & 119   & 7.6      & 25\\
\multicolumn{1}{l}{}                                                                                                                 & \multicolumn{1}{l}{}                                               & \multicolumn{1}{l}{}                                              & \multicolumn{1}{l}{}                                            & 140   & 5.9      & 23\\
\hline

\multicolumn{1}{l}{\multirow{2}{*}{Simons Array \cite{Arnold:2014qym,Creminelli:2015oda}}} 	&\multicolumn{1}{l}{\multirow{2}{*}{\ \ \ \ \ \ 0.65}}   &\multicolumn{1}{l}{\multirow{2}{*}{\ \ \ \ \ 30}}         &\multicolumn{1}{l}{\multirow{2}{*}{\ \ \ \ \ 3000}}    & 95     & 13.9   & 5.2\\
\multicolumn{1}{l}{}                                                                                                                         & \multicolumn{1}{l}{}                                               & \multicolumn{1}{l}{}                                              & \multicolumn{1}{l}{}                                              & 150   & 11.4   & 3.5\\
\hline

\bottomrule
\end{tabular}
\caption{
Design specifications for experiments considered in our forecast.
We only list frequency channels  ($89 {\rm{GHz}} \sim 160 {\rm{GHz}}$) used in our simulation. 
$f_{\rm{sky}}$ and $\ell_{\rm{min}}$ for space-borne missions are set to 0.7 and 2 respectively.
Note that the constraints may improve if the $\ell$ ranges are further optimized with each
experiment's realistic noise model. For example, PICO constraint improves by 53\% when using 
$\ell$ range of $[30, 3000]$ with our noise estimate in Eq.~\ref{eq:noise_par}.
}
\label{Tab_ExpSpecs}
\end{center}
\end{table*}

\end{appendix}
\bibliographystyle{JHEP}
\bibliography{PBH_CMB.bib}

\end{document}